%% file: 8780.tex
\documentclass{aa}
\usepackage{graphicx}
\usepackage{txfonts}
\usepackage{natbib}
\usepackage{supertabular}
\usepackage{longtable}
\usepackage{ulem}
\begin{document}

\newcommand{\zabs}{z_{\rm abs}} 
\newcommand{\dla}{damped Lyman-$\alpha$}
\newcommand{\DLA}{Damped Lyman-$\alpha$}
\newcommand{\lya}{Ly-$\alpha$}
\newcommand{\lyb}{Ly-$\beta$}
\newcommand{\lyg}{Ly-$\gamma$}
\newcommand{\CI}{\ion{C}{i}}
\newcommand{\CII}{\ion{C}{ii}}
\newcommand{\CIV}{\ion{C}{iv}}
\newcommand{\SiIV}{\ion{Si}{iv}}
\newcommand{\HI}{\ion{H}{i}}
\newcommand{\PII}{\ion{P}{ii}}
\newcommand{\SII}{\ion{S}{ii}}
\newcommand{\SiII}{\ion{Si}{ii}}
\newcommand{\FeII}{\ion{Fe}{ii}}
\newcommand{\ZnII}{\ion{Zn}{ii}}
\newcommand{\NiII}{\ion{Ni}{ii}}
\newcommand{\CrII}{\ion{Cr}{ii}}
\newcommand{\TiII}{\ion{Ti}{ii}}
\newcommand{\MnII}{\ion{Mn}{ii}}
\newcommand{\MgI}{\ion{Mg}{i}}
\newcommand{\MgII}{\ion{Mg}{ii}}
\newcommand{\PbII}{\ion{Pb}{ii}}
\newcommand{\CuII}{\ion{Cu}{ii}}
\newcommand{\ArI}{\ion{Ar}{i}}
\newcommand{\NI}{\ion{N}{i}}
\newcommand{\OI}{\ion{O}{i}}
\newcommand{\s}{$\mathcal{S}$}    
\ULforem
\title{Molecular hydrogen in high-redshift Damped Lyman-$\alpha$ systems: The
VLT/UVES database\thanks{Based on data gathered at the European Southern
Observatory (ESO) using the Ultraviolet and Visual Echelle
Spectrograph (UVES) installed at the Very Large Telescope (VLT), Unit-2
(Kueyen), on Cerro Paranal, Chile.}}

\titlerunning{Molecular hydrogen in high-$z$ DLAs: The VLT/UVES database}

\author{P. Noterdaeme\inst{1,2}
   \and C. Ledoux\inst{1}
   \and P. Petitjean\inst{2}
   \and R. Srianand\inst{3}}

\institute{European Southern Observatory, Alonso de C\'ordova 3107, Casilla
19001, Vitacura, Santiago 19, Chile\\
\email{pnoterda@eso.org, cledoux@eso.org}
\and
UPMC Paris 06, Institut d'Astrophysique de Paris - CNRS, 98bis Boulevard Arago, F-75014, Paris, France\\
\email{petitjean@iap.fr}
\and
Inter-University Centre for Astronomy and Astrophysics, Post Bag 4, Ganesh
Khind, Pune 411\,007, India\\
\email{anand@iucaa.ernet.in}}

\date{Received; Accepted}

\abstract{}
{
We present the current status of ongoing searches for molecular
hydrogen in high-redshift ($1.8<\zabs \le 4.2$) Damped Lyman-$\alpha$
systems (DLAs) capitalising on observations performed with the ESO
Very Large Telescope (VLT) Ultraviolet and Visual Echelle Spectrograph
(UVES).
}
{
We identify 77 DLAs/strong sub-DLAs, with $\log N(\HI)\ge 20$ and
$\zabs >1.8$, which have data that include redshifted H$_2$ Lyman
and/or Werner-band absorption lines. This sample of H\,{\sc i}, H$_2$
and metal line measurements, performed in an homogeneous manner, is
more than twice as large as our previous sample (Ledoux et al. 2003)
considering every system in which searches for H$_2$ could be
completed so far, including all non-detections.
}
{
H$_2$ is detected in thirteen of the systems, which have molecular
fractions of values between $f\simeq 5\times 10^{-7}$ and $f\simeq
0.1$, where $f=2N($H$_2)/(2N($H$_2)+N(\HI))$. Upper limits are
measured for the remaining 64 systems with detection limits of
typically $\log N($H$_2)\sim 14.3$, corresponding to $\log f<-5$. We
find that about 35\% of the DLAs with metallicities relative to solar
[X/H$]\ge -1.3$ (i.e., 1/20$^{\rm th}$ solar), with X~=~Zn, S or Si,
have molecular fractions $\log f>-4.5$, while H$_2$ is detected --
regardless of the molecular fraction -- in $\sim 50$\% of them. In
contrast, only about 4\% of the [X/H$]<-1.3$ DLAs have $\log f>-4.5$.
We show that the presence of H$_2$ does not strongly depend on the
total neutral hydrogen column density, although the probability of
finding $\log f>-4.5$ is higher for $\log N(\HI)\ge 20.8$ than below
this limit (19\% and 7\% respectively). The overall H$_2$ detection
rate in $\log N(\HI)\ge20$ DLAs is found to be about 16\% (10\%
considering only $\log f>-4.5$ detections) after correction for a
slight bias towards large $N(\HI)$. There is a strong preference for
H$_2$-bearing DLAs to have significant depletion factors,
[X/Fe$]>0.4$. In addition, all H$_2$-bearing DLAs have column
densities of iron into dust grains larger than $\log N($Fe$)_{\rm
dust}\sim 14.7$, and about 40\% of the DLAs above this limit have
detected H$_2$ lines with $\log f>-4.5$. This demonstrates the
importance of dust in governing the detectability of H$_2$ in DLAs.
Our extended sample supports neither the redshift evolution of the
detection fraction of H$_2$-bearing DLAs nor that of the molecular
fraction in systems with H$_2$ detections over the redshift range
$1.8<\zabs\le 3$.
}
{
}

\maketitle
\section{Introduction}

Damped Lyman-$\alpha$ systems (DLAs) were discovered in the seventies
\citep[e.g., ][]{Lowrance72, Beaver72, Carswell75, Wright79} and
identified afterwards as redshifted damping absorptions from large
column densities of neutral atomic hydrogen \citep[][]{Smith79}.

Numerous DLAs, with $N(\HI)\ge 2\times 10^{20}$\,atoms~cm$^{-2}$, have
been discovered through large dedicated surveys \citep[e.g.,
][]{Wolfe86} and more recently thanks to the huge number of quasar
spectra available from the Sloan Digital Sky Survey
\citep{Prochaska05}. Because DLAs contain most of the neutral hydrogen
available for star formation in the Universe \citep{Wolfe86,
Lanzetta91} and are associated with numerous metal absorption lines,
they probably arise in the interstellar medium of protogalaxies,
progenitors of present-day galaxies \citep[see, e.g., ][]{Wolfe00,
Haehnelt00, Wolfe05}. Our understanding of DLAs is mainly based on the
study of low-ionisation metal absorptions \citep[e.g.,
][]{Prochaska02} but also high-ionisation species \citep{Lu96,
Wolfe00, Fox07a, Fox07b} and, in a few cases, molecular absorptions
\citep[e.g., ][]{Ledoux03}. The latter are not conspicuous however in
contrast to what is seen in the Galaxy and, for a long time, only the
DLA towards Q\,0528$-$2505 was known to contain H$_2$ molecules
\citep{Levshakov85}. H$_2$-bearing DLAs are nevertheless crucial to
understand the nature of DLAs because molecular hydrogen is an
important species to derive the physical conditions in the gas
\citep[see, e.g., ][]{Tumlinson02, Reimers03, Hirashita05, Srianand05,
Cui05, Noterdaeme07}.

The first systematic search for molecular hydrogen in high-redshift
($\zabs >1.8$) DLAs was carried out using the Ultraviolet and Visual
Echelle Spectrograph (UVES) at the Very Large Telescope (VLT)
\citep{Ledoux03}. It consisted of a sample of 33 DLAs with H$_2$
detected in eight of them. Molecular fractions were found to lie in
the range $-3.5<\log f<-1$ with
$f=2N($H$_2)/(2N($H$_2)+N(\HI))$. Upper limits of typically
$N($H$_2)\sim 2\times 10^{14}$~cm$^{-2}$ (corresponding to $\log
f<-5$) were measured in the other systems. More recently, we noted a
correlation between the presence of molecular hydrogen and the
metallicity of high-redshift DLAs \citep{Petitjean06}. High molecular
fractions ($\log f>-4$) were found in about 40\% of the
high-metallicity DLAs ([X/H$]\ge 1/20^{\rm th}$ solar) whilst only
$\sim 5$\% of the [X/H$]<-1.3$ DLAs have $\log f>-4$. Other papers by
our group focused on specific detections. We presented the analysis of
three systems with low molecular fractions, i.e., $\log f<-4$, one of
them having a low metallicity \citep{Noterdaeme07lf}, and the
H$_2$-bearing DLA with, to date, the highest redshift, at $\zabs
=4.224$ towards Q\,1441$+$2737 \citep{Ledoux06b}. We note that a
possible detection of H$_2$ in a DLA towards a Gamma-ray Burst (GRB)
afterglow has been reported recently \citep{Fynbo06}. However, the
origin of DLAs at the GRB host-galaxy redshift is very likely to be
different from those observed in QSO spectra \citep[e.g.,
][]{Jakobsson06, Prochaska07a}.

We present here the whole sample of UVES high-redshift QSO-DLAs for
which the wavelength range where H$_2$ lines are redshifted is covered
by the available spectra. This sample is more than twice as large as
in our previous study \citep{Ledoux03}. We present the observations
and the UVES DLA sample in Sect.~2 and provide comments on individual
absorbers in Sect.~3. We discuss the overall population in Sect.~4 and
results in Sects.~\ref{mol_f} to \ref{z}. We conclude in
Sect.~\ref{conclusion}.

\section{Observations and sample}

All the quasars in our sample were observed with the Ultraviolet and
Visual Echelle Spectrograph \citep[UVES, ][]{Dekker00} mounted on the
ESO VLT-UT\,2 (Kueyen) 8.2~m telescope on Cerro Paranal, Chile. We
have used our UVES database to build up a sample of 77 DLAs/strong
sub-DLAs along 65 lines of sight, hereafter called sample \s. The
systems were selected as having $N(\HI)\ge 10^{20}$~cm$^{-2}$, and
redshifts $\zabs >1.8$ so that at least part of the wavelength range
into which H$_2$ absorption lines are expected to be redshifted, is
covered. Systems for which the flux at the wavelengths of all
redshifted H$_2$ lines is zero as a consequence of the presence of an
intervening Lyman-limit system located at a higher redshift were
rejected. Systems for which the \lya\ forest is so crowded that no
meaningful upper limit on $N$(H$_2$) could be derived were also
excluded from the final sample.

Sample \s\ comprises a total of 68 DLAs ($\log N(\HI) \ge 20.3$) and
nine strong sub-DLAs ($20\le\log N(\HI)<20.3$) according to the common
definition of DLAs \citep[$\log N(\HI)\ge 20.3$; ][]{Wolfe86}. Our
lower limit on $N$(\HI) for sub-DLAs is only slightly lower than the
classical definition of DLAs ensuring that these absorbers are mostly
neutral and share the same physical nature as classical DLAs
\citep{Viegas95}.

Most of the systems in sample \s\ (53 out of 77) come from the sample
of \citet{Ledoux06a}. This sample is mainly drawn from the follow-up
of the Large Bright QSO Survey \citep[LBQS, ][]{Wolfe95} and has been
observed between 2000 and 2004 in the course of our systematic search
for molecular hydrogen at $\zabs >1.8$ \citep[see also][]{Petitjean00,
Ledoux03}. This comprises 46 bona-fide DLAs and seven strong sub-DLAs.

In addition to this sample, we are considering UVES data for 13
absorbers (11 DLAs and two strong sub-DLAs) from the Hamburg-ESO DLA
survey \citep{Smette05}, seven DLAs from the CORALS survey
\citep{Akerman05}, and four DLAs mainly from our own observing
runs. Among the latter systems, the $\zabs =3.692$ and 3.774 DLAs
towards Q\,0131$+$0345 and the $\zabs =2.659$ DLA towards
Q\,0642$-$5038 were observed in visitor mode on September 17-20 2004,
under Prog. ID~073.A-0071 (PI: Ledoux). On the other hand, the $\zabs
=4.203$ DLA towards Q\,0951$-$0450 was observed in service mode on
January 26-27 and February 19-21, 2004, under Prog. ID~072.A-0558 (PI:
Vladilo).

We have reduced all the data including those retrieved from the ESO
archive in an homogeneous manner using the UVES pipeline
\citep{Ballester00}, which is available as a dedicated package of the
ESO MIDAS data reduction system. The main characteristics of the
pipeline are to perform a robust inter-order background subtraction
for master flat-fields and science frames and an optimal extraction of
the object signal subtracting the sky spectrum and rejecting cosmic
rays simultaneously. The wavelength scale of the reduced spectra was
converted to vacuum-heliocentric values. Each spectrum, corresponding
to different instrument settings, was rebinned to a constant
wavelength step. No further rebinning was performed during subsequent
data analysis. Individual exposures were then weighted, scaled and
combined altogether.

Total neutral hydrogen column densities have been measured from
Voigt-profile fitting of the damped Lyman-$\alpha$ and/or
Lyman-$\beta$ lines (see \citealt{Ledoux06a}; Smette et al., in
prep). Most of the metallicity and depletion measurements are from
\citet{Ledoux06a} except when more recently re-measured (see details
in footnote of Table~\ref{sumtable}). Apart from the DLAs towards
Q\,0131$+$0345 and Q\,0951$-$0450 (for which abundances were taken
from \citet{Prochaska07b} because UVES observations did not allow a
sufficient number of metal lines to be covered), we (re)determined all
metal column densities by the homogeneous fitting of Voigt-profiles to
non-saturated absorption lines. This includes the systems from the
Hamburg-ESO and CORALS DLA surveys. Metallicities were determined
using the zinc abundance (when \ZnII\ is detected), or from the
abundances of sulphur or silicon. No ionisation correction has been
applied. Zinc and sulphur are known to be little depleted into dust
grains. Silicon, in turn, is probably mildly depleted and has been
used only in cases where Zn and S are not detected. The results are
summarised in Table~\ref{sumtable}. Abundances are given relative to
solar, i.e., [X/H$]\equiv\log N($X$)/N($H$)-\log$~(X/H)$_\odot$. Solar
abundances as listed in \citet{Morton03}, based on meteoritic data
from \citet{Grevesse02}, were adopted.

\section{Comments on individual systems \label{comments}}

\par\noindent
{\sl Q\,0000$-$2619, $\zabs=3.390$:}
\citet{Levshakov00, Levshakov01} reported a tentative detection of
H$_2$ in this system. However, they detected only two weak absorption
features in the Lyman-$\alpha$ forest, identified as H$_2$\,W2-0\,Q1
and L4-0\,R1. The probability that these features are actually due to
intervening Lyman-$\alpha$ absorbers is high and we consider the
derived $J=1$ column density as an upper limit only. In any case, this
system has a very low molecular fraction of $\log f<-6.87$.
\par\noindent
{\sl Q\,0013$-$0029, $\zabs=1.973$:}
The detection of H$_2$ in this system has been reported for the first
time by \citet{Ge97}. \citet{Petitjean02} showed that this system is
actually the blend of a DLA ($\log N(\HI)=20.83$) and a sub-DLA ($\log
N(\HI)\le 19.43$) separated by $\sim$500~km\,s$^{-1}$. H$_2$ is
present in both systems, with four components detected up to $J=5$.
\par\noindent
{\sl Q\,0027$-$1836, $\zabs=2.402$:}
This system has the lowest metallicity ([X/H]~$=-1.63$) of systems in which H$_2$ molecules have been detected. H$_2$ is detected in a single component up to rotational level $J=5$ and possibly 6. Thanks to the high data quality, we detected an increase in the Doppler parameter $b$, from low to high rotational levels, for the first time at high redshift \citep{Noterdaeme07lf}.
\par\noindent
{\sl Q\,0347$-$3819, $\zabs=3.025$:}
The detection of H$_2$ in this system was first reported by
\citet{Levshakov02} using UVES commissioning data. Subsequently
\citet{Ledoux03} analysed higher quality spectra of the system. Weak
molecular absorption is observed in a single component up to $J=4$.
\par\noindent
{\sl Q\,0405$-$4418, $\zabs=2.595$:}
Molecular hydrogen is detected up to $J=3$ in a single component
\citep[see][]{Ledoux03, Srianand05}. There is some indication that the
relative strengths of the $J=2$ and 3 lines require $b\ga
1.5$~km\,s$^{-1}$, while the $J=0$ and 1 lines are consistent with
$b=1.3$~km\,s$^{-1}$.
\par\noindent
{\sl Q\,0528$-$2505, $\zabs=2.811$:}
This is the first DLA system in which H$_2$ was detected
\citep{Levshakov85}. The detection of H$_2$ was readdressed in more
detail by \citet{Foltz88}, \citet{Srianand98} and
\citet{Srianand05}. The single H$_2$ component seen in a low
resolution CASPEC spectrum \citep{Srianand98} is resolved into two
components in a UVES spectrum \citep{Srianand05}. Rotational levels up
to $J=5$ are detected.
\par\noindent
{\sl Q\,0551$-$3638, $\zabs=1.962$:}
This system has the highest metallicity in the UVES DLA sample with
[Zn/H$]=-0.35$. H$_2$ is clearly detected in two well-resolved
features separated by about 55~km\,s$^{-1}$ \citep{Ledoux02}. The
first feature is narrow and weak, but detected up to $J=3$. The second
feature is broader and stronger and has been fitted using two
components at precisely the same redshift as the detected \CI\ lines.
\par\noindent
{\sl Q\,0642$-$5038, $\zabs=2.659$:}
H$_2$ in this system is detected in a single strong component. This is
a new detection and details will be presented in a forthcoming paper
(Ledoux et al., in prep.).
\par\noindent
{\sl Q\,0841$+$1256, $\zabs=2.375$:}
\citet{Petitjean00} detected two weak absorption features
corresponding to H$_2$\,L4-0\,R0 and L2-0\,R0 at the same redshift. As
for the DLA towards Q\,0000$-$2619, we consider the tentative
measurement of the $J=0$ column density as an upper limit only.
\par\noindent
{\sl Q\,1232$+$0815, $\zabs=2.338$:}
A single H$_2$ component is seen in rotational levels $J=0$ to 5
\citep{Ge01,Srianand05}. This is the only DLA in which deuterated
molecular hydrogen (HD) has been detected until now
\citep{Varshalovich01}.
\par\noindent
{\sl Q\,1441$+$2737, $\zabs=4.224$:}
Because of the high redshift of this system (actually the highest
amongst known H$_2$-bearing DLAs), the Lyman-$\alpha$ forest is quite
dense in the wavelength region in which H$_2$ lines are
redshifted. However, H$_2$ could be detected and studied in a large
number of transitions thanks to the high spectral resolution of the
data. Four rotational levels are detected in no less than three
velocity components \citep{Ledoux06b}.
\par\noindent
{\sl Q\,1444$+$0126, $\zabs=2.087$:}
Molecular hydrogen is found in two components of this sub-DLA in $J\le
3$ rotational levels \citep{Ledoux03}.
\par\noindent
{\sl Q\,1451$+$1223, $\zabs=2.469$:}
Because of the Lyman-break from the DLA at $\zabs=3.171$, the expected
positions of only two lines (H$_2$\,L0-0\,R0 and L0-0\,R1) fall in a
region of non-zero flux. The high upper limit we derive is not
strongly constraining and the molecular fraction is probably well
below $\log f=-4.5$.
\par\noindent
{\sl Q\,2318$-$1107, $\zabs=1.989$:}
Weak H$_2$ absorption features from $J=0$ to 2 are unambiguously
detected in a single component \citep{Noterdaeme07lf}.
\par\noindent
{\sl Q\,2343$+$1232, $\zabs=2.431$:}
Thanks to the high quality of the spectrum, H$_2$ absorption lines
(from $J=0$ and 1) are unambiguously detected although they are
extremely weak ($N($H$_2)\sim 5\times10^{13}$~cm$^{-2}$). We measure
the lowest molecular fraction yet measured for a DLA, $\log f=-6.41$
\citep{Petitjean06, Noterdaeme07lf}.
\par\noindent
{\sl Q\,2348$-$0108, $\zabs=2.426$:}
No less than seven H$_2$ components are detected in this system
\citep{Petitjean06, Noterdaeme07}. The H$_2$ profile is complex with
three strong and four weak components spread over about
250~km\,s$^{-1}$, making H$_2$ lines from different transitions
overlap. However, the large number of observed transitions and the
good data quality allow an accurate measurement of column density for
different rotational levels.

\section{Overall population \label{overall_pop}}

In this Section, we use the results summarised in Table~\ref{sumtable}
to derive the properties of the H$_2$ gas in the global DLA
population. We list in this table, the name of the quasar in column 1,
the emission and absorption redshifts in columns 2 and 3, the total
\HI\ column density in column 4, the metallicity relative to solar
[X/H] and the depletion factor [X/Fe] in columns 5 and 6, with X given
in column 7, the velocity spread of the low-ionisation metal line
profiles \citep[see][]{Ledoux06a} in column 8, the total molecular
hydrogen column densities in the $J=0$ and 1 rotational levels in
columns 9 and 10, the mean molecular fraction in column 11, and the
references for the measurements in column 12. Upper limits on the
molecular fraction were calculated to be the sum of the upper limits
on the first two rotational levels ($J=0$ and 1) and are given at the
3\,$\sigma$ significance level. The thirteen firm detections of H$_2$
constitute what we will hereafter call sub-sample \s$_{\rm H2}$. Ten
of them have molecular fractions higher than all upper limits, i.e.,
$\log f>-4.5$, and will constitute sub-sample \s$_{\rm HF}$ (standing
for high molecular fractions). We note that obviously \s$_{\rm
HF}\subset$~\s$_{\rm H2}\subset$~\s.

\begin{figure}[!ht]
 \begin{center}
\begin{tabular}{c} 
 \includegraphics[clip=,width=\hsize,bb=23 390 600 750]{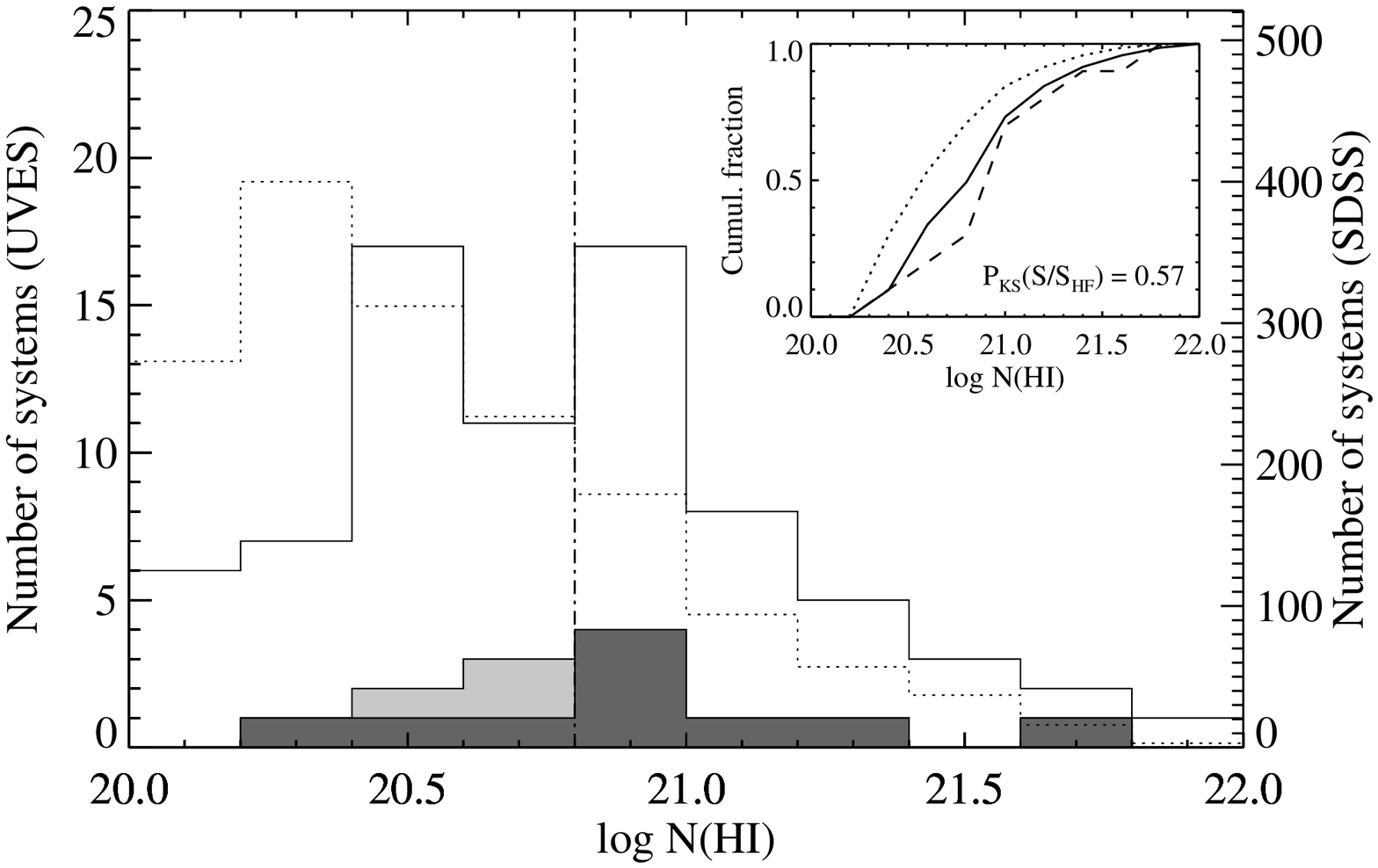}\\
 \includegraphics[clip=,width=\hsize,bb=23 390 600 750]{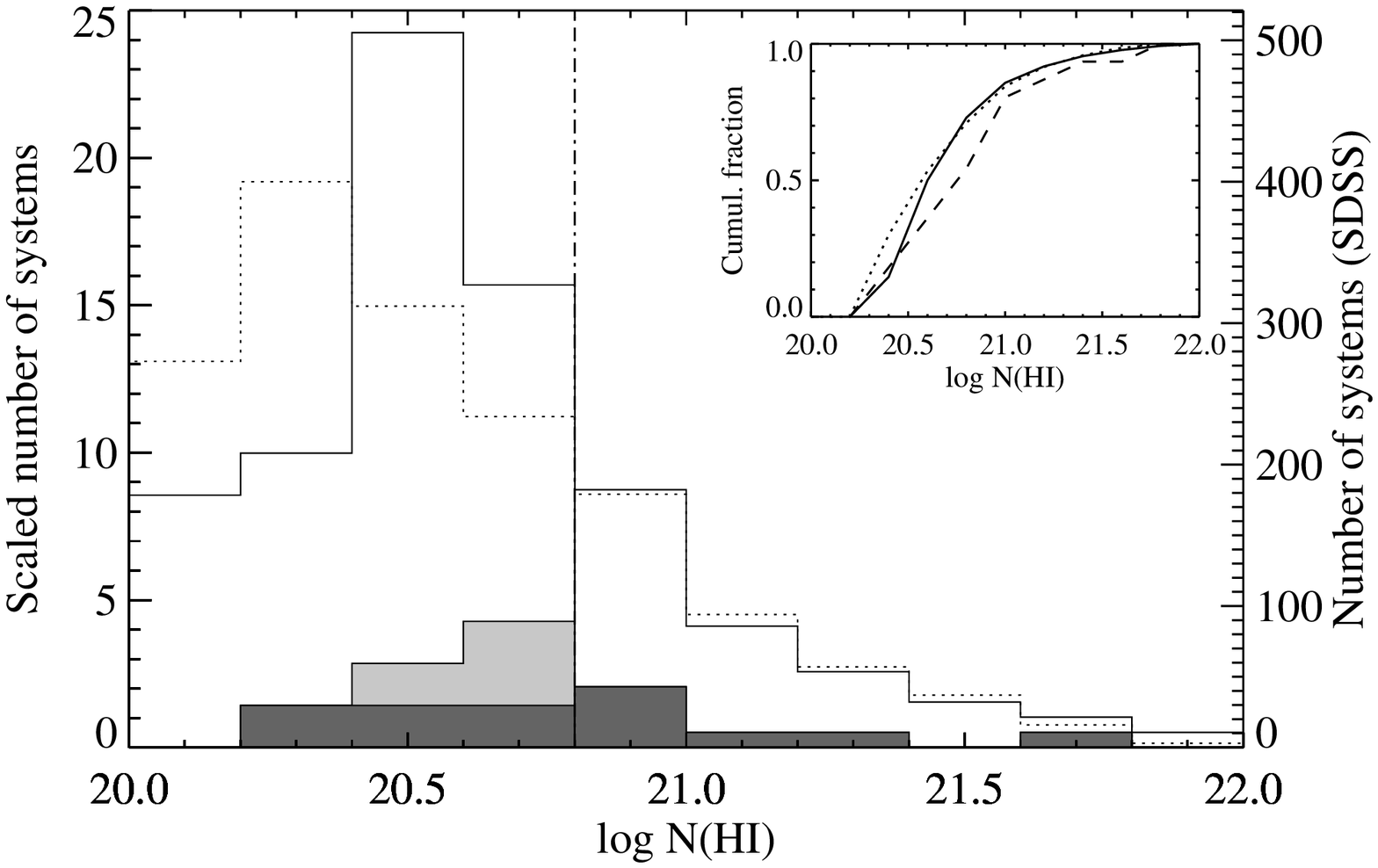}\\
\end{tabular}
 \caption{\label{hist_lognhi} {\sl Top panel}:
Neutral hydrogen column density distributions of DLAs in the overall
UVES sample \s\ (solid), the sub-sample of H$_2$-bearing systems
\s$_{\rm H2}$ (grey), and that of the systems with $\log f>-4.5$
(\s$_{\rm HF}$; dark grey). The distribution from the SDSS-DR5 sample
\citep[dotted; ][]{Prochaska05} is represented with a different
scaling (right axis) so that the area of both histograms are the same.
The inset shows the cumulative distributions from sample \s\ (solid
line), sub-sample \s$_{\rm HF}$ (dashed), and the SDSS-DR5 sample
(dotted). {\sl Bottom panel}: The above $\log N(\HI)$ distributions
are corrected for the bias in favour of large $N(\HI)$ column
densities (see text) by scaling the numbers separately for $20\le\log
N(\HI)<20.8$ and $\log N(\HI)\ge 20.8$ to match the SDSS-DR5
distribution.}
 \end{center}
\end{figure}

The upper panel of Fig.~\ref{hist_lognhi} gives the distribution of
$N(\HI)$ from sample \s, sub-samples \s$_{\rm H2}$ and \s$_{\rm HF}$
to be compared to the reference distribution derived from the SDSS-DR5
DLA sample scaled to the number of systems in the UVES sample
\citep{Prochaska05}. It is apparent that the distribution from the
UVES sample is slightly biased compared to the scaled SDSS
distribution\footnote{Note that the SDSS $N$(\HI) distribution itself
may be biased, however it is currently the best available
representation of the DLA population.} in favour of large $N(\HI)$
DLAs (i.e., with $\log N(\HI)\ge 20.8$). We can however correct for
this bias. We divide each DLA sample into two halves, above and below
$\log N(\HI)=20.8$. We scale the number of systems in each bin of the
UVES distribution by the ratio of the number of systems in the
(scaled) SDSS and UVES samples, above (by a ratio of $\sim 0.5$) and
below (by $\sim 1.4$) $\log N(\HI)=20.8$. The result of this scaling
is shown in the bottom panel of Fig.~\ref{hist_lognhi}. In the inset,
the cumulative distributions are shown in order to compare the two
populations. It is apparent that the scaling has corrected for the
above-mentioned bias.

The overall H$_2$ detection rate in $\log N(\HI)\ge 20$ systems is
found to be about 16\% (10\% considering only $\log f>-4.5$
detections) after correction for the above-mentioned bias. There is no
H$_2$ detection for $\log N(\HI)<20.2$. These numbers, however, may
well be affected by small number statistics. H$_2$ is in fact detected
in a sub-DLA component ($\log N(\HI)\sim 19.4$) of the $\zabs =1.973$
system towards Q\,0013$-$0029 \citep[see][]{Petitjean02}. The
corresponding molecular fraction is larger than $10^{-4}$. In any
case, additional observations are required to investigate the
molecular content of sub-DLAs. Apart, possibly, from the first bin
($\log N(\HI)<20.2$), there is no strong dependence of the presence of
molecules on the neutral hydrogen column density. Large molecular
fractions are seen over the range $20.2\le\log N(\HI)< 21.8$.

The double-sided Kolmogorov-Smirnov test yields a probability $P_{\rm
KS}= 0.57$ that sample \s\ and sub-sample \s$_{\rm HF}$ are drawn from
the same parent population.


However, molecular hydrogen with $\log f>-4.5$ is detected in
respectively 9\% and 19\% of the systems in the two sub-samples:
$20.2\le\log N(\HI)<20.8$ (3/35) and $\log N(\HI)\ge 20.8$ (7/36),
which implies that there is probably a tendency for H$_2$ to be seen
more often in large column density systems. We note that the fraction
is 7\% (3/41) for the range $20.0\le\log N(\HI)<20.8$. We also note
that we are considering here the total neutral hydrogen column
densities. The actual \HI-clouds corresponding to the H$_2$
absorptions probably have smaller $N(\HI)$. This is in fact observed
for the DLA at $\zabs =1.973$ towards Q\,0013$-$0029, for which
partial deblending of the \lya\ absorption is possible.

\begin{figure}[!ht]
 \begin{center}
\begin{tabular}{c}
 \includegraphics[clip=,width=\hsize,bb=28 390 600 750]{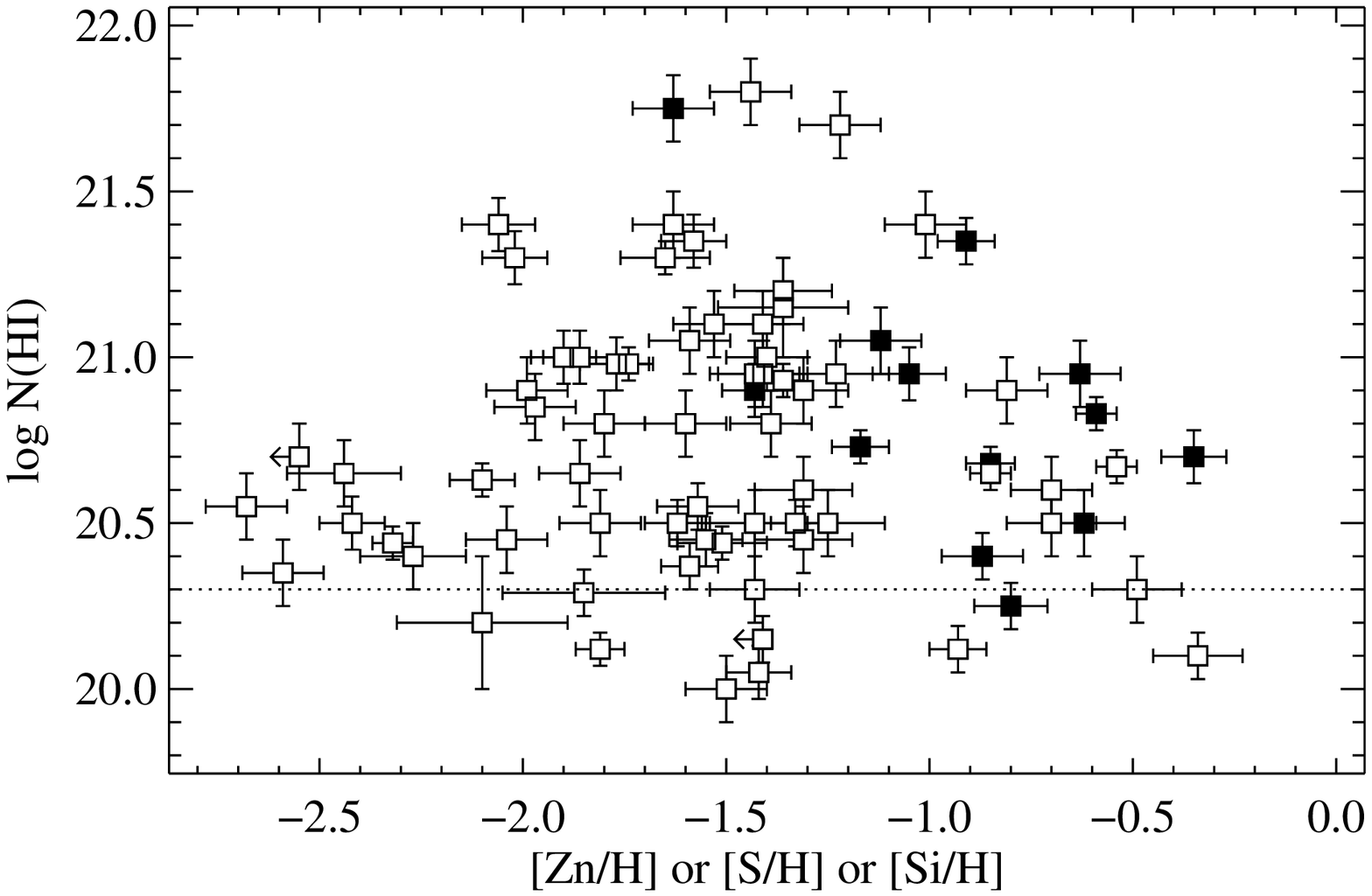}\\
 \includegraphics[clip=,width=\hsize,bb=28 390 600 750]{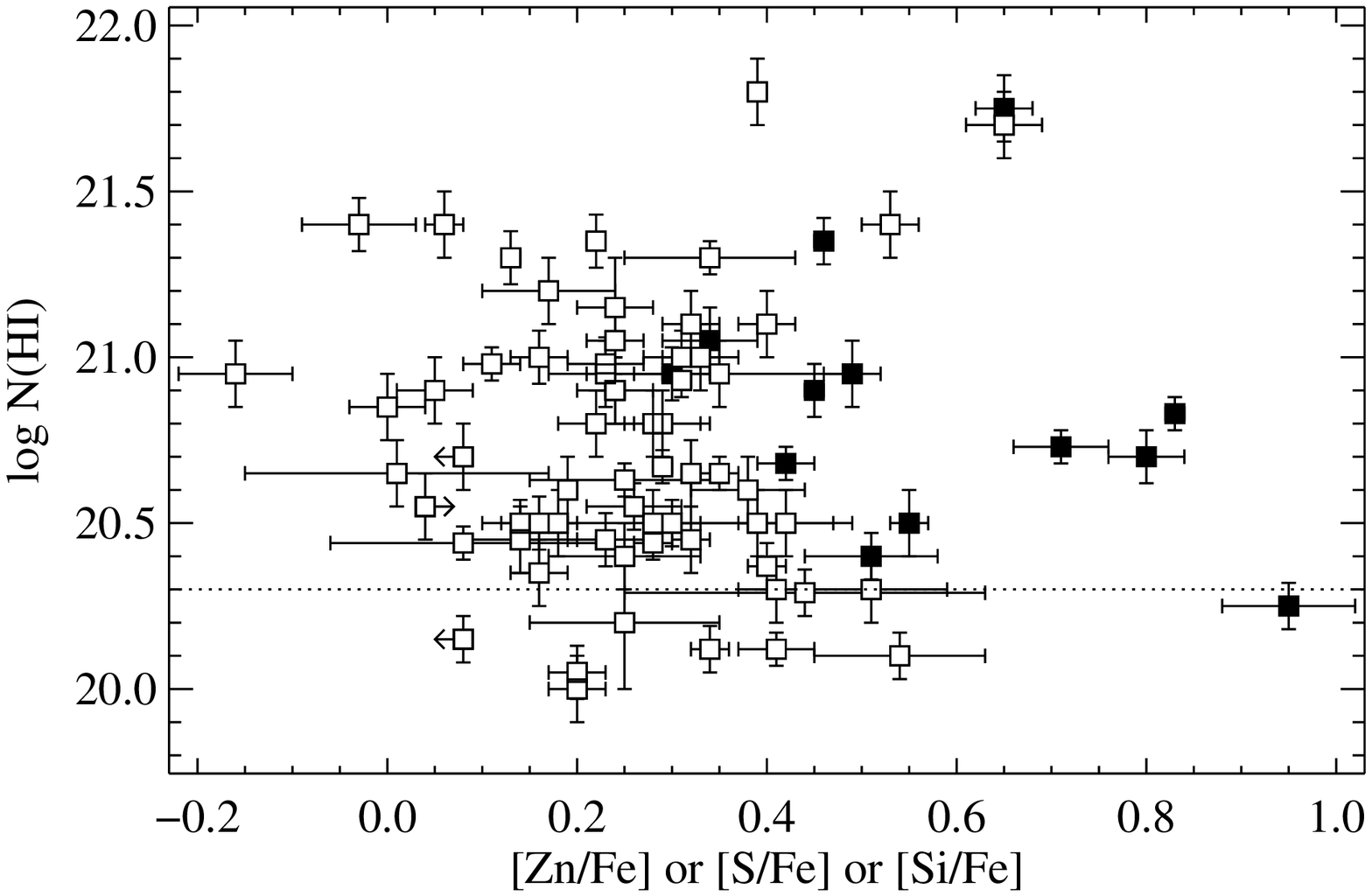}\\
\end{tabular}
 \caption{\label{lognhi_xhxfe}
Logarithm of the total neutral hydrogen column density in the UVES
sample versus metallicity (top panel) or depletion factor (bottom
panel). Filled squares indicate systems in which H$_2$ is
detected. The lack of systems with both a high metallicity and a large
\HI\ column density is seen in all DLA samples. There is no
correlation between $N(\HI)$ and [X/H] or [X/Fe].}
 \end{center}
\end{figure}

The fact that the sample is biased in favour of large \HI\ column
density DLAs could be a problem for the analysis of the overall DLA
population if any correlation exists between $\log N(\HI)$ and the
metallicity ([X/H]) and/or the depletion of metals onto dust grains
([X/Fe]). This is not the case however: Figure~\ref{lognhi_xhxfe}
shows the neutral hydrogen column density as a function of metallicity
(top panel) and depletion factor (bottom panel). It is apparent that
there is no correlation between [X/H] or [X/Fe] and $\log N(\HI)$. The
Kendall's rank correlation coefficient is as small as 0.02
(resp. $-0.01$) for [X/H] (resp. [X/Fe]) vs. $\log N(\HI)$. The lack
of systems with both a high metallicity and large $N(\HI)$ is a
feature common to all DLA samples. It can be explained either by a
lack of large \HI\ column densities because of the transition of \HI\
into H$_2$ \citep{Schaye01} and/or by the fact that any quasar located
behind such an absorber would remain undetected because of the induced
extinction \citep[e.g.,][]{Boisse98,Vladilo05}. Such systems could
also be associated with regions with low projected cross-sections.
Therefore, although the lack of high-metallicity systems with large
$N(\HI)$, together with the bias towards large $N(\HI)$ in our sample,
could introduce a small bias towards low metallicities, the absence of
the above-described correlations implies that any bias will have
little influence on the properties based on the metallicities and
depletion factors discussed in Sect.~\ref{sect_dust}.

\begin{figure}[!ht]
 \begin{center}
 \includegraphics[clip=,width=\hsize,bb=23 390 600 750]{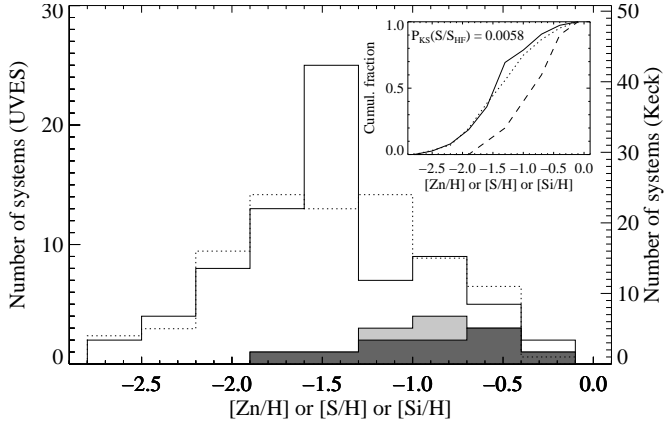}\\
 \caption{\label{hist_xh}
The metallicity distribution of DLAs in the overall UVES sample (\s ;
solid) is compared to that from the Keck sample \citep[dotted;
][]{Prochaska07b}, adequately scaled (right axis) so that the area of
both histograms are the same. The distributions are similar.
Distributions from sub-samples \s$_{\rm H2}$ (grey) and \s$_{\rm HF}$
(dark grey) are also shown. It is clear from this and the
Kolmogorov-Smirnov test probability (inset) that the distributions
from H$_2$-bearing systems are skewed towards high metallicities.}
 \end{center}
\end{figure}

In Fig.~\ref{hist_xh}, we compare the metallicity distribution of DLAs
in the UVES sample to that of DLAs in the Keck sample
\citep[HIRES+ESI, ][]{Prochaska07b}. Although a difference can be seen
in the range $-1.6<[$X/H$]<-1$, partly due to the bias towards large
$N$(\HI), no systematic bias is observed for the UVES sample, apart
from any also affecting the Keck sample, as shown by the cumulative
distributions (inset). We note that the observed distributions are
similar in shape to the distribution derived using simulations by
\citet[][their fig.~3]{Hou05}. The metallicity distribution for
H$_2$-bearing DLAs is also plotted. It is apparent from the figure
that the distribution of H$_2$-bearing systems is strongly skewed
towards high metallicities (see Sect.~\ref{sect_dust}).

\section{Molecular fraction \label{mol_f}}

\begin{figure}[!Ht]
\begin{center}
 \includegraphics[clip=,width=0.95\hsize,bb=8 -50 600 770]{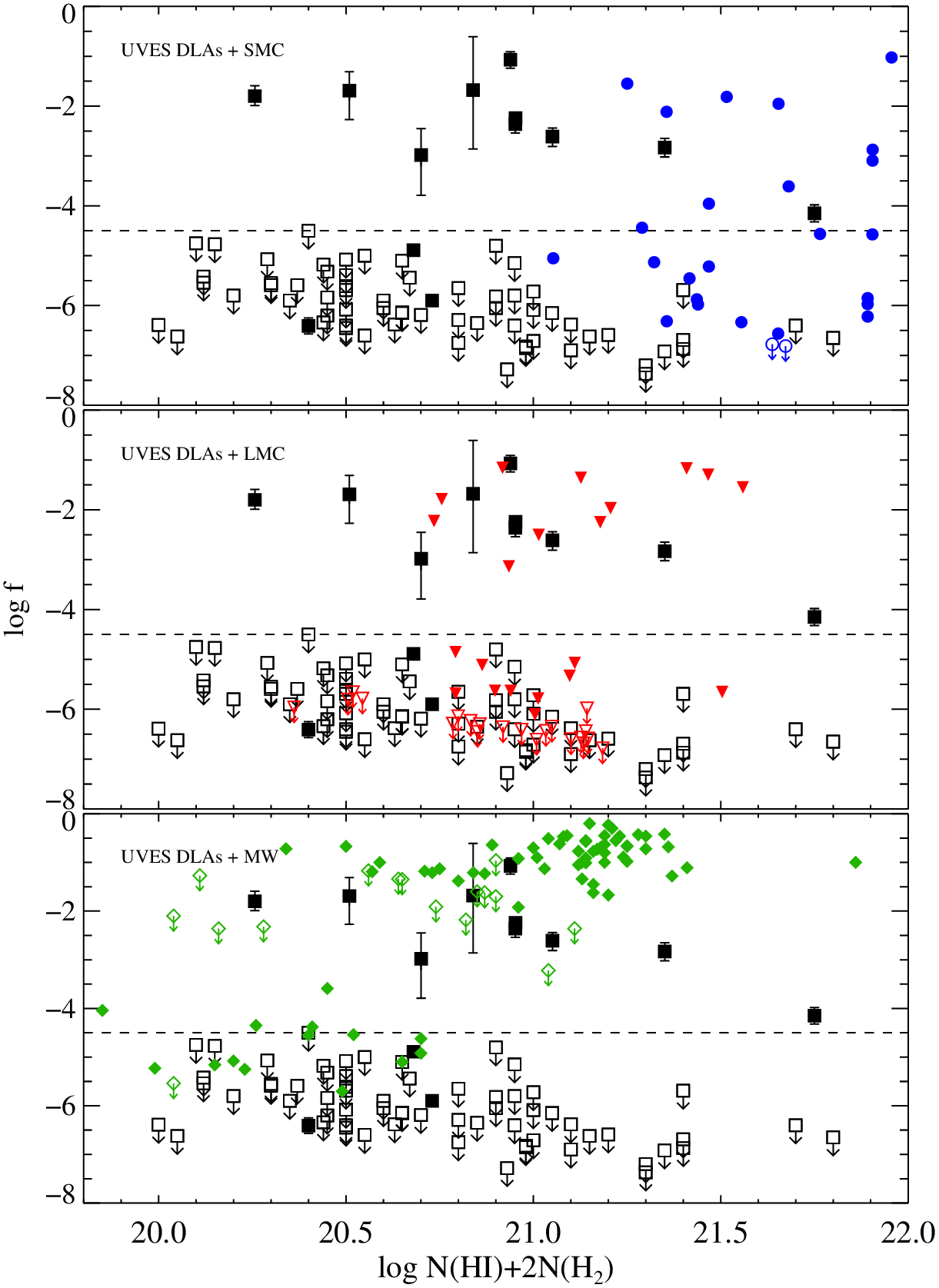}
\caption{\label{logf_lognhih2}
Molecular fraction versus total hydrogen ($\HI+$H$_2$) column density
for sample \s\ (the filled squares correspond to H$_2$ detections,
i.e., sub-sample \s$_{\rm H2}$), lines of sight through the SMC
\citep[blue dots, top panel; ][]{Tumlinson02}, the LMC \citep[red
triangles, middle panel; ][]{Tumlinson02} and the Galactic disk
\citep[green diamonds, bottom panel; ][]{Savage77}. Upper limits are
marked by unfilled symbols with downward arrows.}
 \end{center}
\end{figure}

As can be seen in Fig.~\ref{logf_lognhih2}, there is no sharp
transition in the molecular fraction of DLAs at any total hydrogen
($\HI+$H$_2$) column density $N$(H), in contrast to what is observed
at $\log N$(H)~=~20.7 in the Galactic disk \citep[][bottom panel of
Fig.~\ref{logf_lognhih2}]{Savage77} or at $\log N$(H)~=~20.4 along
high-latitude Galactic lines of sight \citep[][]{Gillmon06}. We note
also that no sharp transition from \HI\ to H$_2$ is observed in the
Magellanic clouds (upper and middle panels of
Fig.~\ref{logf_lognhih2}), although SMC lines of sight probe larger
gas column densities. Such transitions are expected to occur at $\log
N$(H)~$>21.3$ in the LMC and $\log N$(H)~$>22$ in the SMC
\citep{Tumlinson02}. The molecular fractions of DLAs are similar to
values measured for the Magellanic clouds, but are lower than
measurements for the Galactic disk. The absence of a sharp \HI/H$_2$
transition does not contradict the predictions by \citet{Schaye01}
because the covering factor of large $N($H$_2)$ systems is expected to
be small \citep{Zwaan06}. Moreover, the induced extinction could be
large (see below) and such systems may have so far been missed. We
should compile a representative sample of strong DLAs ($\log N(\HI)>
21.5$) in order to address this point.

\section{The importance of dust \label{sect_dust}}

\begin{figure}[!ht]
 \begin{center}
 \includegraphics[clip=,width=\hsize,bb=28 390 600 750]{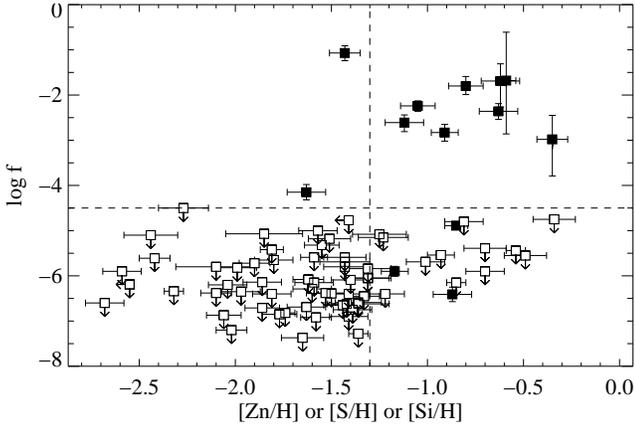}
 \caption{\label{logf_xh}
Logarithm of the molecular fraction, $f=2N($H$_2)/(2N($H$_2)+N(\HI)$,
versus metallicity, [X/H$]=\log N($X$)/N($H$)-\log ({\rm X/H})_\odot$
with X$=$Zn, S or Si. Filled squares indicate systems in which H$_2$
is detected. The horizontal (resp. vertical) dashed line indicates the
separation between what we somewhat arbitrarily call high and low
molecular fractions, $\log f=-4.5$ (resp. high and low metallicities,
[X/H$]=-1.3$).}
 \end{center}
\end{figure}

Figure~\ref{logf_xh} is an extension of fig.~3 from
\citet{Petitjean06} to the whole UVES sample, with about 2.5 times
more systems with [X/H$]<-1.3$. It can be seen from this figure that
about 35\% of the [X/H$]\ge -1.3$ (1/20$^{\rm th}$ solar) systems have
molecular fractions $\log f>-4.5$, while H$_2$ is detected --
regardless of the molecular fraction -- in $\sim 50$\% of them. In
contrast, only about 4\% of the [X/H$]<-1.3$ DLAs have $\log
f>-4.5$. We remind the reader that we use the limit $\log f>-4.5$ to
define systems with high molecular fraction because all of our upper
limits are below this value. The lowest metallicity at which H$_2$ has
been detected to date in a DLA is [Zn/H$]=-1.63$
\citep{Noterdaeme07lf}. The corresponding system, at $\zabs =2.402$
towards Q\,0027$-$1836, has a molecular fraction of $\log f=-4.15$.

\begin{figure}[!ht]
 \begin{center}
 \includegraphics[clip=,width=\hsize,bb=28 390 600 750]{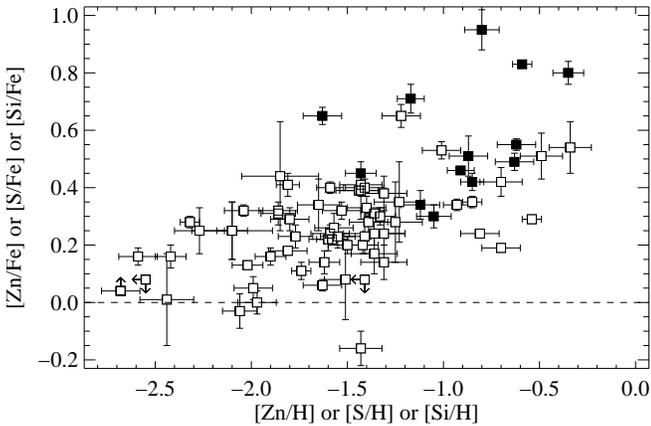}
 \caption{\label{xfe_xh}
A correlation (5.3\,$\sigma$ confidence level) exists in the overall
UVES sample between dust-depletion factor, [X/Fe], and metallicity,
[X/H], with X either Zn, S or Si.}
 \end{center}
\end{figure}

This trend in metallicity supports the idea that the presence of dust
is an important ingredient in the formation of H$_2$. The correlation
between the depletion of metals into dust grains and the metallicity
\citep{Ledoux03} has now been confirmed using a sample more than twice
the size, and at higher confidence level (5.3\,$\sigma$ compared to
4\,$\sigma$) (see Fig.~\ref{xfe_xh}). High-metallicity DLAs are
usually more dusty and, as a consequence, the formation rate of H$_2$
onto dust grains is enhanced, while the photo-destruction rate is
lowered by dust- and self-shielding.

\begin{figure}[!ht]
 \begin{center}
 \includegraphics[clip=,width=\hsize,bb=28 390 600 750]{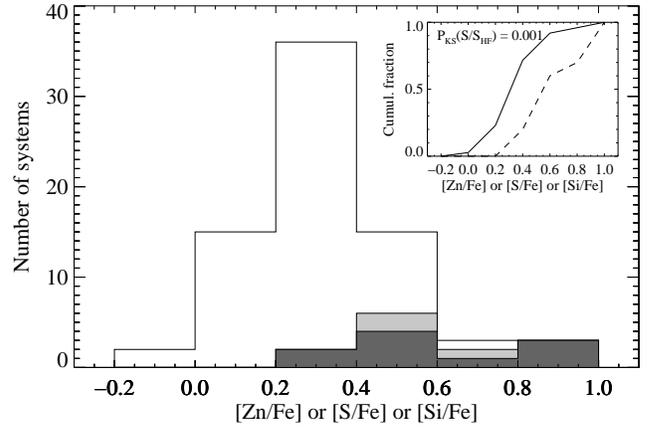}
 \caption{\label{hist_xfe}
Distribution of depletion factors from the overall UVES sample (\s ;
solid), sub-samples \s$_{\rm H2}$ (grey) and \s$_{\rm HF}$ (dark
grey). It is clear from these histograms that the distribution of
depletion factors for H$_2$-bearing systems is different from that of
the overall sample. This shows clearly that H$_2$-bearing DLAs are
more dusty than the rest of the DLAs. The probability that the two
samples are drawn from the same parent population is indeed very
small: $P_{\rm KS}($\s/\s$_{\rm HF})\approx 10^{-3}$.}
 \end{center}
\end{figure}

Figure~\ref{hist_xfe} shows the distribution of depletion factors
(defined as [X/Fe], with X$=$Zn, S or Si) for the different samples of
systems defined in this paper. It is clear from this histogram that
H$_2$ is found in DLAs with the highest depletion factors. The
probability that sample \s\ and sub-sample \s$_{\rm H2}$ are drawn
from the same parent population is very small, i.e., $P_{\rm
KS}\approx 10^{-3}$.

\begin{figure}[!ht]
 \begin{center}
 \includegraphics[clip=,width=\hsize,bb=28 390 600 750]{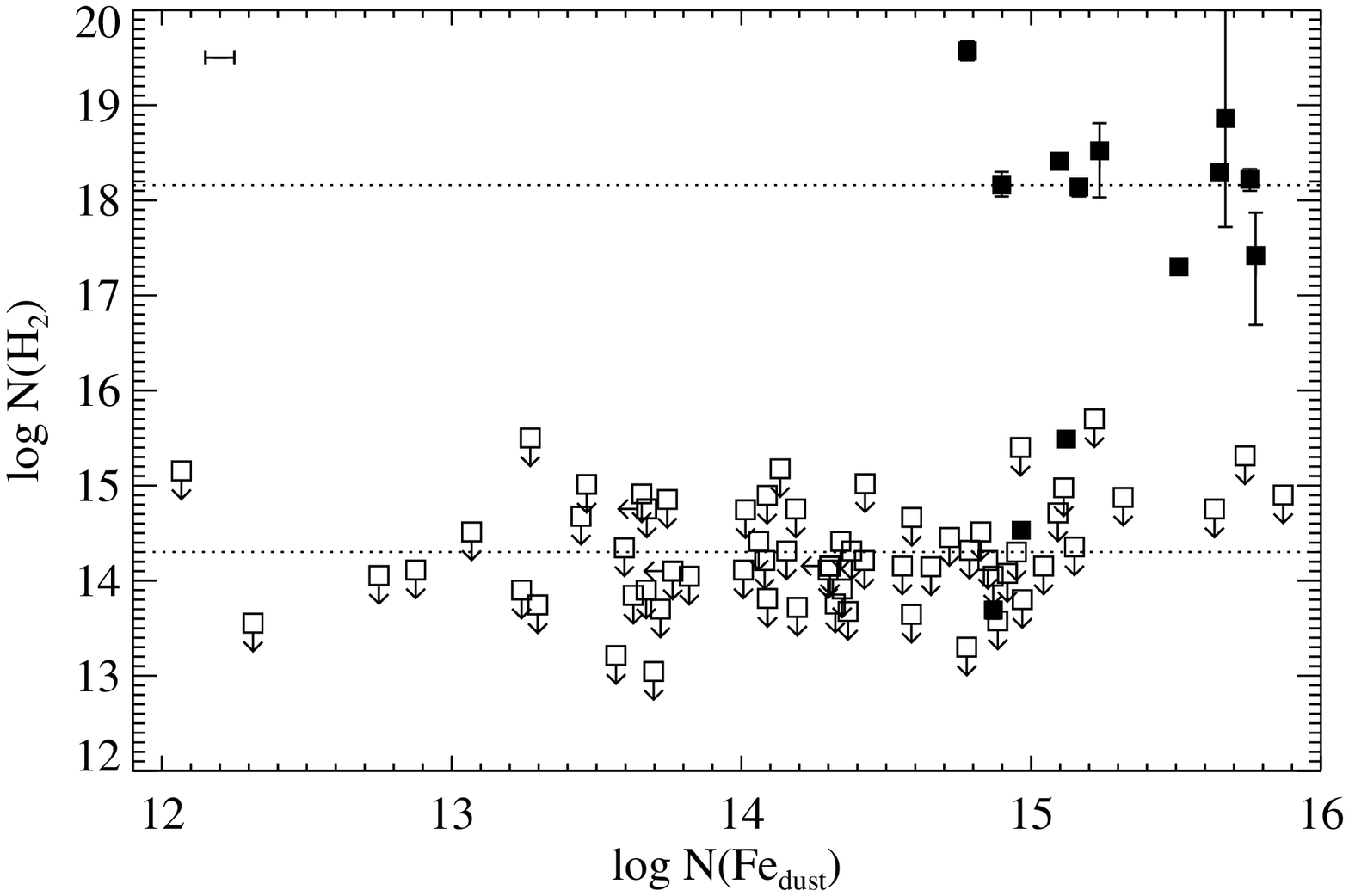}
 \caption{\label{nh2_nfed}
H$_2$ column density, $N($H$_2)$, versus column density of iron into
dust grains, $N($Fe$)_{\rm dust}$. The horizontal dotted line at $\log
N$(H$_2$)~$\simeq$~18.2 (resp. 14.3) shows the median $\log N$(H$_2$)
(resp. the median upper limit on $\log N($H$_2)$) for the
H$_2$-bearing DLAs from sub-sample \s$_{\rm H2}$ (resp. the
H$_2$-undetected systems from sub-sample \s~$-$~\s$_{\rm H2}$). There
is a clear gap of about four orders of magnitude between the two
values. The horizontal bar in the upper-left corner of the plot shows
the typical uncertainty on the measured $\log N$(Fe)$_{\rm dust}$.}
 \end{center}
\end{figure}

To test the influence of the presence of dust on the molecular
hydrogen column density, it is worth comparing $N$(H$_2$) to the
column density of iron into dust, $N($Fe$)_{\rm
dust}=(1-10^{-\left[{\rm X/Fe}\right]})N({\rm X})\left({\rm
Fe/X}\right)_{\rm dla}$ (see Fig.~\ref{nh2_nfed}). \citet{Vladilo06}
showed that absorbers associated with large column densities of iron
in dust, generally induce more extinction (i.e. have larger values of
$A_{\rm V}$). This correlation is consistent with that observed along
Galactic interstellar lines of sight \citep[e.g., ][]{Snow02}. We note
that the above expression of $N$(Fe)$_{\rm dust}$ assumes that X is a
non-refractory element and that the intrinsic [X/Fe] ratio is
solar. \citet{Vladilo02} introduced corrections for zinc depletion and
possible non-solar intrinsic abundances \citep[i.e., (Fe/X$)_{\rm
dla}\ne ($Fe/X$)_\odot$; see also][]{Vladilo06} that we do not apply
here because the corrections are small, and not necessary for our
purposes. This effect is most important for Silicon that can be
depleted into dust grains, and the depletion factor is underestimated
when using [Si/Fe]. The effect is however probably less than 0.3 dex
\citep[see][]{Petitjean02} and should have no consequence on the
characteristics of the whole population. When [X/Fe] is negative,
which happens for three DLAs, $N($Fe$)_{\rm dust}$ cannot be computed
directly. We thus estimated a 3\,$\sigma$ upper limit on $N($Fe$)_{\rm
dust}$ by considering the upper bound provided by the 3\,$\sigma$
error on [X/Fe].

It can be seen on Fig.~\ref{nh2_nfed} that all H$_2$-bearing DLAs have
$N($Fe$)_{\rm dust}>5\times 10^{14}$~cm$^{-2}$, supporting the idea
that dust is an important ingredient in the formation of H$_2$. We
note however that this column density corresponds to a small
extinction (${\log A_{\rm V}\sim -1.5}$; \citealt{Vladilo06}) possibly
explaining the absence of detectable H$_2$ in three DLAs at
$N($Fe$)_{\rm dust}>3\times 10^{15}$~cm$^{-2}$.

An additional reason why H$_2$ is not detected in these systems could
be related to the particle density in the neutral gas being too small.
The equilibrium between formation and destruction of H$_2$ molecules
can be written as $R n n($H$)=R_{\rm diss} n($H$_2)$, where $R_{\rm
diss}$ is the photo-dissociation rate, $R$ is the formation rate, $n$
the particle density, $n($H$)\simeq n$ the proton density and
$n($H$_2)$ the H$_2$ density. By multiplying this expression by the
longitudinal size of the cloud, one can see that the H$_2$ column
density depends linearly on the particle density. Large dust column
densities may occasionally result from large and diffuse clouds where
the particle density is insufficiently large for H$_2$ to form
efficiently enough to be detectable. The column densities of iron in
dust are in fact large ($N($Fe$)_{\rm dust}>3\times
10^{15}$~cm$^{-2}$) in three DLAs without detectable H$_2$ mainly
because of the large neutral hydrogen column densities (i.e. $\log
N(\HI)=21.70$ towards Q\,0458$-$0203; $\log N(\HI)=21.80$ towards
Q\,1157$+$0128; and $\log N(\HI)=21.40$ towards Q\,1209$+$0919).

The fact that H$_2$ is detected in some DLAs at low extinction, could
imply that the weak (due to high depletion) iron component, associated
with the H$_2$-bearing component, is hidden in the overall metal-line
profile \citep[see, e.g., ][]{Ledoux02}. This may bias the
measurements of metallicities and depletion factors, and hence the
measurement of column density of iron into dust grains. In the future,
we propose that this issue is more deeply investigated.

\section{Gas kinematics}

\begin{figure}[!ht]
 \begin{center}
 \includegraphics[clip=,width=0.95\hsize,bb=28 390 600 750]{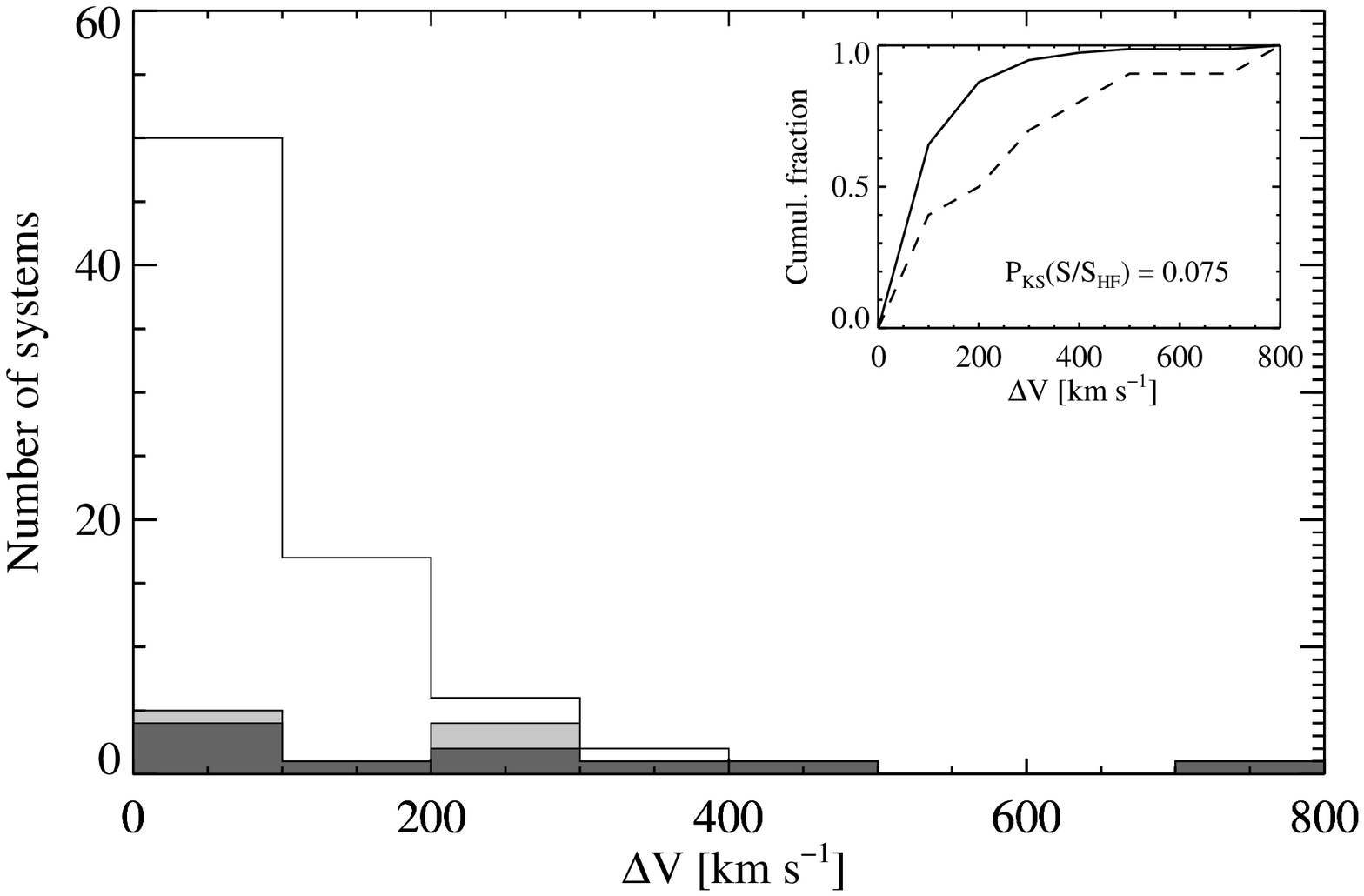}
 \caption{\label{hist_deltav}
Distribution of the velocity spread, $\Delta V$, of low-ionisation
metal line profiles \citep[see][for details on the
technique]{Ledoux06a} from the overall UVES sample (\s, 77 systems,
solid histogram), systems with detected molecular hydrogen (\s$_{\rm
H2}$, 13 systems, grey), and systems with $\log f>-4.5$ (\s$_{\rm
HF}$, ten systems, dark grey). It is apparent from this and the
Kolmogorov-Smirnov test probability (inset) that the likelihood of
finding H$_2$ is higher when $\Delta V$ is larger.}
 \end{center}
\end{figure}

Figure~\ref{hist_deltav} shows the distribution of the velocity widths
of the low-ionisation metal line profiles. These measurements were
completed using the method described in \citet{Ledoux06a}, and many
were previously published in that paper. It is clear from
Fig.~\ref{hist_deltav} that the probability of finding H$_2$ is higher
when the velocity width of the low-ionisation metal line profiles is
larger. Using the double-sided Kolmogorov-Smirnov test, we calculate a
probability $P_{\rm KS}=0.075$ that samples \s\ and \s$_{\rm HF}$ are
derived from the same parent distribution. A natural explanation is
provided by \citet{Ledoux06a} who suggested that higher metallicity
DLAs arise from more massive objects. The amount of molecular gas and
the star-formation rate in these systems could therefore be enhanced
\citep{Hirashita05} naturally explaining the above correlation.
Another explanation could be that outflows can pull out cold and dusty
gas, as well as providing a large number of velocity components that
would increase the probability of finding H$_2$ \citep{Murphy07}.
However, H$_2$ is usually detected in few components (typically one or
two) in the main clumps of the systems, and not in the satellite
components at high velocities that make up most of the width of the
profile. The only exceptions to this are the DLA at $\zabs\simeq2.43$
towards Q\,2348$-$0108 where no less than seven H$_2$ components are
spread over 250~km\,s$^{-1}$ \citep{Noterdaeme07}, and the
$\zabs=1.973$ system towards Q\,0013$-$0029 where two pairs of H$_2$
components are separated by more than 500~km\,s$^{-1}$
\citep{Petitjean02}. The latter could be considered as the blend of
two DLA systems (see Sect.~\ref{comments}). In addition,
\citet{Fox07a} interpreted DLAs where the \CIV\ total line width
exceeds the escape velocity to be associated with outflowing
winds. Using this criterion, there is no indication that H$_2$ is
found preferably in DLAs with outflows. Indeed, considering the 54 DLA
systems common to the present sample and that of \citet{Fox07a}, we
can estimate that H$_2$ is detected in $\sim$16\% (3/19) of the DLAs
with outflows while this percentage is $\sim$22\% (12/54) for the
entire DLA population.

\section{Evolution with redshift \label{z}}

\begin{figure}[!ht]
 \begin{center}
\begin{tabular}{c}
 \includegraphics[clip=,width=0.95\hsize,bb=42 390 600 750]{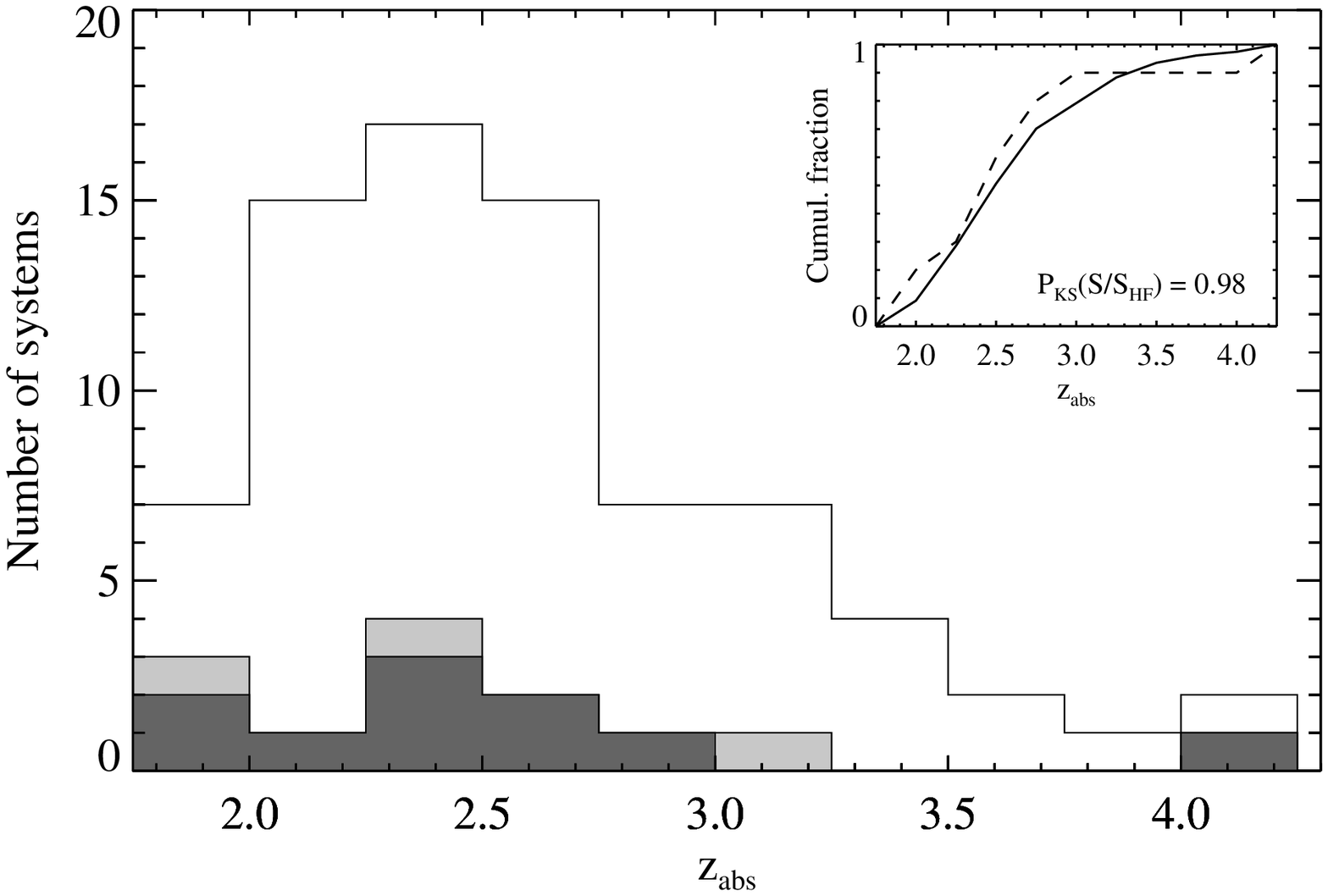}\\
 \includegraphics[clip=,width=0.95\hsize,bb=42 390 600 750]{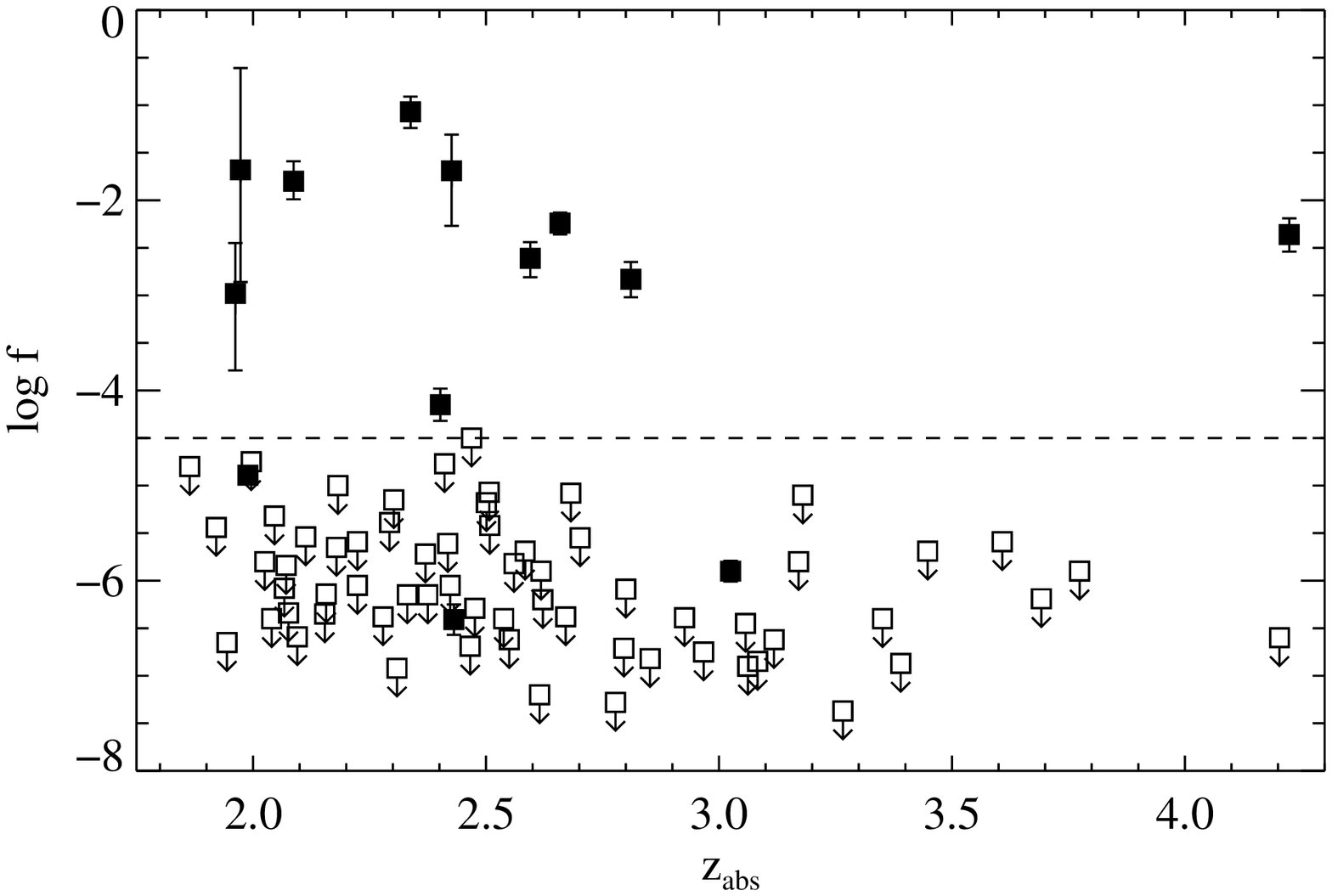}\\
\end{tabular}
 \caption{\label{hist_z} {\sl Top panel}:
Redshift distributions of DLAs in the overall UVES sample (\s, 77
systems, solid histogram), systems with detected molecular hydrogen
(\s$_{\rm H2}$, 13 systems, grey), and systems with $\log f>-4.5$
(\s$_{\rm HF}$, ten systems, dark grey). The inset shows the
cumulative distributions from sample \s\ (solid) and sub-sample
\s$_{\rm HF}$ (dashed). The two distributions are statistically
indistinguishable ($P_{\rm KS}=0.98$). {\sl Bottom panel}: Logarithm
of the molecular fraction, $\log f$, versus absorber redshift. There
is a slight trend for more constraining upper limits at higher
redshift. This is not physical however but due to the higher
signal-to-noise ratios achieved towards the red, together with the
larger number of H$_2$ transitions covered by the spectra.}
 \end{center}
\end{figure}

In Fig.~\ref{hist_z}, we analyse the redshift distribution of DLAs
within the UVES sample. The H$_2$-bearing DLA sub-sample is
statistically indistinguishable from the overall UVES sample. A
double-sided Kolmogorov-Smirnov test shows that sample \s\ and
sub-sample \s$_{\rm H2}$ have a probability $P_{\rm KS}=0.94$ to be
drawn from the same parent population. There is no indication of any
evolution of the fraction of H$_2$-bearing DLAs with
redshift. Molecular hydrogen is more difficult to detect at the lowest
redshifts observable using UVES ($z_{\rm abs}\la 2$) because fewer and
weaker transitions are then covered by the UVES spectra and lines are
located in the bluest part of the spectra where the signal-to-noise
ratios are lower. While this could imply a bias in the detection rate
of H$_2$ in systems with low molecular fractions, high molecular
fractions (i.e., $\log f>-4.5$) are detected anyway because the
corresponding column densities are always well above the detection
limit. When considering only high molecular fractions, the same
Kolmogorov-Smirnov test yields $P_{\rm KS}$(\s/\s$_{\rm
HF})=0.98$. Contrary to previous tentative evidence \citep{Curran04},
there is also no evolution of the molecular fraction in systems with
detected H$_2$ (see Fig.~\ref{hist_z}).

However, because of small number statistics for systems at $z>3$, it
is still unclear whether there is no really evolution in the sense of
fewer H$_2$ detections at higher redshift. Therefore, we restrict our
claim of no evolution with redshift of the detection fraction and the
molecular fraction to the range $1.8<\zabs\le 3$. On the other hand,
large molecular fractions are observed in the local Universe and
significant evolutions of both the fraction of H$_2$-detected systems
and the molecular fractions are expected at lower redshifts. It is
then of prime importance to cover the redshift interval $0\le\zabs \le
1.8$. This will soon be possible using the Cosmic Origins Spectrograph
to be installed onboard the Hubble Space Telescope.

\section{Conclusion \label{conclusion}}

We present results of the largest survey of molecular hydrogen in
high-redshift ($1.8<\zabs \le 4.2$) DLAs compiled to date using high
signal-to-noise ratio, high spectral-resolution VLT-UVES data. We
analyse data for 77 DLAs/strong sub-DLAs with $N(\HI)\ge
10^{20}$~cm$^{-2}$, a dataset more than twice as large as that studied
previously by \citet{Ledoux03}. From the thirteen high-redshift
H$_2$-bearing DLAs known to date, nine have been discovered by our
group. Due to the superb quality of the Ultraviolet and Visual Echelle
Spectrograph, we are able to detect unambiguously the H$_2$ absorption
features, measure accurate column densities, and in the cases of
non-detections derive stringent upper limits.

A double-sided Kolmogorov-Smirnov test shows that the ten
H$_2$-bearing systems with $\log f>-4.5$ (which is our conservative
completeness limit) have \HI\ column densities that are compatible
with those of the overall DLA population. This may be due however to
small number statistics. There is evidence (see
Sect.~\ref{overall_pop}) that the probability of finding large
molecular fractions is higher in DLAs with large $N(\HI)$, as observed
in the LMC. About $7$\% of the systems with $\log N(\HI)<20.8$ have
$\log f>-4.5$ while $\sim 19$\% of the systems with larger \HI\ column
densities have similar molecular fractions. There is no sharp
transition in the molecular fraction of DLAs at any of the measured
total hydrogen ($\HI+$H$_2$) column density.

It is surprising to see that most of the DLAs have very low molecular
fractions, i.e., $\log f \la -6$. This is much smaller than observed
along lines of sight in the Galactic disk \citep{Savage77}. However,
\citet{Wakker06} measured similar molecular fractions towards
high-latitude Galactic lines of sight.

We confirm that a good criterion to find H$_2$-bearing DLAs is to
select high-metallicity systems. Indeed, 35\% of the systems with
[X/H$]\ge -1.3$ have $\log f>-4.5$ whilst this is the case for only
4\% of those with metallicities lower than that. Since there is a
correlation between metallicity and depletion factor, the latter being
defined as [X/Fe] with X$=$Zn, S or Si, H$_2$ is found in DLAs also
having the highest depletion factors. Therefore, clouds with large
molecular fractions are expected to be dusty and clumpy
\citep[e.g.,][]{Hirashita03}. They can be missed because of the
associated extinction and/or because of their small cross-section.

The presence of H$_2$ is closely related to the dust column
density. Indeed, all detections pertain to systems where the column
density of iron into dust grains is larger than $5\times
10^{14}$~cm$^{-2}$, and about 40\% of these systems have detectable
H$_2$. This shows that the presence of dust is an important ingredient
in the formation of H$_2$ in DLAs. The low molecular fractions
measured for most of the DLAs are probably a consequence of low
abundances of metals and dust.

We show that the probability of finding H$_2$ increases with
increasing velocity width of low-ionisation metal line profiles. The
correlation between velocity width and metallicity \citep{Ledoux06a},
if interpreted as a mass-metallicity relation, provides a natural
explanation in which H$_2$-bearing DLA systems are preferably
associated with massive objects where star-formation is enhanced.
There is no evidence of systematic outflows in H$_2$-bearing DLAs
neither from the H$_2$ profiles nor from the study of the associated
high-ionisation phase.

From the comparison between H$_2$-bearing systems and the overall UVES
sample, we show that there is no evolution with redshift of the
fraction of H$_2$-bearing DLAs nor of the molecular fraction in
systems with detected H$_2$ over the range $1.8<\zabs\le 3$. This,
compared to the large amounts of H$_2$ observed in the local Universe,
suggests that a significant increase of the molecular fraction in DLAs
could take place at redshifts $\zabs\le 1.8$. Ultraviolet
observations from space are therefore needed to observe the H$_2$
Lyman and Werner bands in low and intermediate redshift DLAs.

Increasing the sample of H$_2$-bearing DLAs is required to cover a
large range in column density and derive the H$_2$ frequency
distribution, $f(N($H$_2),${\sl X}) -- where {\sl X} is the absorption
distance -- in a way similar to studies completed for \HI\
\citep{Lanzetta91}. This would assess whether the steep slope observed
in $f(N(\HI),${\sl X}) at large column densities \citep[e.g.,
][]{Prochaska05} is due to the conversion \HI$\rightarrow$H$_2$
\citep{Schaye01} rather than to a magnitude-dependent bias from dust
obscuration of the background quasars \citep[e.g., ][]{Boisse98,
Vladilo05, Smette05}. Both effects can explain the apparent lack of
DLAs of both high metallicity and large \HI\ column density, as
observed in all DLA samples.

While the exact nature of DLAs is still an open debate, H$_2$-bearing
DLAs provide interesting probes of the physical conditions close to
star-forming regions. About a decade ago, only one H$_2$-bearing DLA
was known. With the criteria revealed by the present survey, it will
be possible to select these systems more efficiently and their numbers
should increase rapidly. This opens up the exciting prospect of the
detailed study of the ISM in distant galaxies.

\acknowledgement{We thank an anonymous referee and the language editor
for useful comments that improved the paper. PN is supported by a PhD
studentship from ESO. PPJ and RS gratefully acknowledge support from
the Indo-French Centre for the Promotion of Advanced Research (Centre
Franco-Indien pour la Promotion de la Recherche Avanc\'ee) under
contract No. 3004-3.
}

\bibliographystyle{aa}
\bibliography{8780bib}

\input 8780_tab

\end{document}

%% file: 8780_tab.tex

\longtab{1}{
\begin{longtable}{c c c c c c c c c c c l}
\caption{\label{sumtable} Molecular and metal contents of UVES DLAs/sub-DLAs at $1.8<\zabs\le4.2$.}\\
\hline \hline
  Quasar        & $z_{\rm em}$	& $z_{\rm abs}$	& $\log N(\ion{H}{i})$  &	[X/H$]$		&    [X/Fe$]$	        & X  &$\Delta V$     & \multicolumn{2}{c}{$\log N$(H$_2$)}       	&	$\log f$	           & Refs. \\        	
      B1950     &               &               &                       &                       &                       &    &[km\,s$^{-1}$] & J~=~0                   & J~=~1                 &                                  &        \\
\hline
\endfirsthead
\caption{continued.}\\
\hline \hline
  Quasar        & $z_{\rm em}$	& $z_{\rm abs}$	& $\log N(\ion{H}{i})$  &	[X/H$]$		&    [X/Fe$]$	        & X  &$\Delta V$     & \multicolumn{2}{c}{$\log N$(H$_2$)}       	&	$\log f$	           & Refs. \\        	
       B1950    &               &               &                       &                       &                       &    &[km\,s$^{-1}$] & J~=~0                   & J~=~1                 &                                  &        \\
\hline
\endhead
\hline
\endfoot
Q\,0000$-$2619 & 	4.11	&	3.390	&	$21.40\pm0.08$	&	$-2.06\pm0.09$	&  $-0.03\pm0.06$	& Zn & 33  & $<13.9$           	        &	$\le 13.8$         &	$<-6.87$		   & $a,b$            \\  
Q\,0010$-$0012 & 	2.15	&	2.025	&	$20.95\pm0.10$	&	$-1.43\pm0.11$	&  $-0.16\pm0.06$	& Zn & 32  & $<14.4$         	        &	$<14.5$		&	$<-5.80$		   & $c$            \\	
Q\,0013$-$0029 & 	2.09	&	1.973	&	$20.83\pm0.05$	&	$-0.59\pm0.05$	&   $0.83\pm0.01$       & Zn & 720 & \multicolumn{2}{c}{$18.86^{+1.14}_{-1.14}$}	&	$-1.68^{+1.07}_{-1.18}$	   & ${c,d,e}$        \\	
Q\,0027$-$1836 & 	2.56	&	2.402	&	$21.75\pm0.10$	&	$-1.63\pm0.10$	&   $0.65\pm0.03$	& Zn & 44  & \multicolumn{2}{c}{$17.30^{+0.07}_{-0.07}$}	&	$-4.15^{+0.17}_{-0.17}$	   & $f,g$            \\  
Q\,0039$-$3354 & 	2.48	&	2.224	&	$20.60\pm0.10$	&	$-1.31\pm0.12$	&   $0.38\pm0.06$	& Si & 122 & $<13.8$    	      	&	$<13.9$		&	$<-6.05$		   & $g,h$            \\  
Q\,0049$-$2820 & 	2.26	&	2.071	&	$20.45\pm0.10$	&	$-1.31\pm0.12$	&   $0.14\pm0.06$	& Si & 51  & $<13.8$    	      	&	$<14.0$		&	$<-5.84$		   & $g,h$            \\  
Q\,0058$-$2914 & 	3.09	&	2.671	&	$21.10\pm0.10$	&	$-1.53\pm0.10$	&   $0.32\pm0.03$	& Zn & 34  & $<13.7$        	        &	$<14.2$		&	$<-6.38$		   & $c$            \\	
Q\,0100$+$1300 & 	2.69	&	2.309	&	$21.35\pm0.08$	&	$-1.58\pm0.08$	&   $0.22\pm0.01$	& Zn & 37  & $<13.5$    	      	&	$<13.9$		&	$<-6.92$		   & $h$            \\	
Q\,0102$-$1902 & 	3.04	&	2.370	&	$21.00\pm0.08$	&	$-1.90\pm0.08$	&   $0.16\pm0.03$	& S  & 17  & $<14.2$         	        &	$<14.8$		&	$<-5.72$		   & $c$            \\	
Q\,0102$-$1902 & 	3.04 	&	2.926	&	$20.00\pm0.10$	&	$-1.50\pm0.10$	&   $0.20\pm0.03$	& Si & 146 & $<12.8$    	      	&	$<13.0$		&	$<-6.39$		   & $h$            \\	
Q\,0112$-$3030 & 	2.99	&	2.418	&	$20.50\pm0.08$	&	$-2.42\pm0.08$	&   $0.16\pm0.04$	& Si & 31  & $<14.1$         	        &	$<14.3$		&	$<-5.61$		   & $c$            \\	
Q\,0112$-$3030 & 	2.99	&	2.702	&	$20.30\pm0.10$	&	$-0.49\pm0.11$	&   $0.51\pm0.08$	& Si & 218 & $<14.1$          	        &	$<14.0$	        &	$<-5.55$		   & $c$            \\  
Q\,0112$+$0259 & 	2.81	&	2.423	&	$20.90\pm0.10$	&	$-1.31\pm0.11$	&   $0.24\pm0.04$	& S  & 112 & $<14.1$        	        &	$<14.2$		&	$<-6.05$		   & $c$            \\	
Q\,0131$+$0345 &        4.15    &       3.692   &       $20.70\pm0.10$  &       $<-2.55$        &   $<0.08$             & S  & 39  & $<13.7$                    &       $<13.9$         &       $<-6.19$                   & ${h,\dagger}$  \\  
Q\,0131$+$0345 &        4.15    &       3.774   &	$20.60\pm0.10$  &       $-0.70\pm0.10$  &   $0.19$              & Si & 128 & $<14.0$                    &       $<14.0$         &       $<-5.90$                   & ${h,\ast}$     \\  
Q\,0135$-$2722 & 	3.21	&	2.800	&	$21.00\pm0.10$	&	$-1.40\pm0.10$	&   $0.33\pm0.04$	& S  & 65  & $<14.3$        	        &	$<14.1$		&	$<-6.09$		   & $c$            \\	
Q\,0216$+$0803 & 	2.99	&	2.293	&	$20.50\pm0.10$	&	$-0.70\pm0.11$	&   $0.42\pm0.05$	& Zn & 104 & $<14.3$          	        &	$<14.5$	     	&	$<-5.39$		   & $i$            \\  
Q\,0242$-$2917 & 	3.27	&	2.560	&	$20.90\pm0.10$	&	$-1.99\pm0.10$	&   $0.05\pm0.04$	& S  & 70  & $<14.5$    	      	&	$<14.2$		&	$<-5.82$		   & $g,h$            \\  
Q\,0254$-$4025 & 	2.28	&	2.046	&	$20.45\pm0.08$	&	$-1.55\pm0.09$	&   $0.23\pm0.04$	& S  & 37  & $<14.2$    	      	&	$<14.6$		&	$<-5.32$		   & $g,h$            \\  
Q\,0300$-$3152 & 	2.37	&	2.179	&	$20.80\pm0.10$	&	$-1.80\pm0.11$	&   $0.29\pm0.04$	& S  & 41  & $<14.5$    	      	&	$<14.4$		&	$<-5.65$		   & $g,h$            \\  
Q\,0331$-$4506 & 	2.67	&	2.411	&	$20.15\pm0.07$	&       $<-1.41$    	&   $<0.08$        	& Si & 31  & $<14.6$	    	      	&	$<14.8$		&	$<-4.77$		   & $g,h$            \\  
Q\,0335$-$1213 &        3.44    &       3.180   &       $20.65\pm0.10$  &       $-2.44\pm0.14$  &   $0.01\pm0.16$       & Si & 15  & $<14.9$                    &       $<14.8$         &       $<-5.10$                   & $h,j$            \\  
Q\,0336$-$0143 & 	3.20	&	3.062	&	$21.10\pm0.10$	&	$-1.41\pm0.10$	&   $0.40\pm0.03$	& Si & 67  & $<13.5$    	      	&	$<13.5$		&	$<-6.90$		   & $h$            \\	
Q\,0347$-$3819 & 	3.22	&	3.025	&	$20.73\pm0.05$	&	$-1.17\pm0.07$	&   $0.71\pm0.05$	& Zn & 93  & \multicolumn{2}{c}{$14.53^{+0.06}_{-0.06}$}	&	$-5.90^{+0.11}_{-0.11}$	   & ${c,k}$        \\	
Q\,0405$-$4418 & 	3.02	&	2.550	&	$21.15\pm0.15$	&	$-1.36\pm0.16$	&   $0.24\pm0.04$	& Zn & 165 & $<13.9$        	        &	$<13.6$		&	$<-6.62$		   & $c$            \\	
Q\,0405$-$4418 & 	3.02	&	2.595	&	$21.05\pm0.10$	&	$-1.12\pm0.10$	&   $0.34\pm0.05$	& Zn & 79  & \multicolumn{2}{c}{$18.14^{+0.07}_{-0.10}$}	&	$-2.61^{+0.17}_{-0.20}$	   & ${c,l}$        \\	
Q\,0405$-$4418 & 	3.02	&	2.622	&	$20.45\pm0.10$	&	$-2.04\pm0.10$	&   $0.32\pm0.02$	& Si & 182 & $<13.3$        	        &	$<13.7$		&	$<-6.20$		   & $c$            \\	
Q\,0421$-$2624 & 	2.28	&	2.157	&	$20.65\pm0.10$	&	$-1.86\pm0.10$	&   $0.32\pm0.01$	& Si & 47  & $<13.7$    	      	&	$<13.9$		&	$<-6.14$		   & $g,h$            \\  
Q\,0425$-$5214 & 	2.25	&	2.224	&	$20.30\pm0.10$	&	$-1.43\pm0.11$	&   $0.41\pm0.04$	& S  & 40  & $<13.9$	    	      	&	$<14.1$		&	$<-5.59$		   & $g,h$            \\  
Q\,0432$-$4401 &        2.65    &       2.302   &       $20.95\pm0.10$  &       $-1.23\pm0.13$  &   $0.35\pm0.14$       & Si & 88  & $<15.1$                    &       $<15.1$         &       $<-5.15$                   & $h,j$            \\  
Q\,0450$-$1310 & 	2.25	&	2.067	&	$20.50\pm0.07$	&	$-1.62\pm0.08$	&   $0.14\pm0.04$	& S  & 148 & $<13.5$    	      	&	$<13.9$		&	$<-6.08$		   & $h$            \\	
Q\,0458$-$0203 & 	2.29	&	2.040	&	$21.70\pm0.10$	&	$-1.22\pm0.10$	&   $0.65\pm0.04$	& Zn & 88  & $<14.6$          	        &	$<14.6$  	&	$<-6.40$		   & $i,m$            \\  
Q\,0528$-$2505 & 	2.77	&	2.811	&	$21.35\pm0.07$	&	$-0.91\pm0.07$	&   $0.46\pm0.01$	& Zn & 304 & \multicolumn{2}{c}{$18.22^{+0.11}_{-0.12}$}	&	$-2.83^{+0.18}_{-0.19}$	   & ${b,l,n,o,p}$  \\	
Q\,0551$-$3638 & 	2.32	&	1.962	&	$20.70\pm0.08$	&	$-0.35\pm0.08$	&   $0.80\pm0.04$	& Zn & 468 & \multicolumn{2}{c}{$17.42^{+0.45}_{-0.73}$}	&	$-2.98^{+0.53}_{-0.81}$	   & $c,q$            \\	
Q\,0642$-$5038 & 	3.09	&	2.659	&	$20.95\pm0.08$	&	$-1.05\pm0.09$  &   $0.30\pm0.04$	& Zn & 99  & \multicolumn{2}{c}{$18.41^{+0.03}_{-0.04}$}	&	$-2.24^{+0.11}_{-0.12}$	   & $h$            \\  
Q\,0841$+$1256 & 	2.50	&	2.375	&	$21.05\pm0.10$	&	$-1.59\pm0.10$	&   $0.24\pm0.03$	& Zn & 37  & $\le 14.56$                &	$<14.0$		&	$<-5.98$		   & $b$            \\	
Q\,0841$+$1256 & 	2.50	&	2.476	&	$20.80\pm0.10$	&	$-1.60\pm0.10$	&   $0.22\pm0.04$	& Zn & 30  & $<13.7$         	        &	$<13.9$		&	$<-6.29$		   & $b$          \\	
Q\,0913$+$0714 & 	2.78	&	2.618	&	$20.35\pm0.10$	&	$-2.59\pm0.10$	&   $0.16\pm0.03$	& Si & 22  & $<13.7$    	      	&	$<13.8$		&	$<-5.90$		   & $h$            \\	
Q\,0933$-$3319 &        2.91    &       2.682   &       $20.50\pm0.10$  &       $-1.27\pm0.14$  &   $0.28\pm0.14$       & Si & 312 & $<14.9$                    &       $<14.4$         &       $<-5.08$                   & $h,j$            \\  
Q\,0951$-$0450 &        4.37    &       4.203   &       $20.55\pm0.10$  &       $-2.68\pm0.10$  &   $>0.04$             & Si & 15  & $<13.2$                     &       $<13.3$         &       $<-6.60$                   & ${h,\dagger}$  \\  
Q\,1036$-$2257 & 	3.13	&	2.778	&	$20.93\pm0.05$	&	$-1.36\pm0.05$	&   $0.31\pm0.01$	& S  & 80  & $<13.0$    	      	&	$<13.0$		&	$<-7.28$		   & $h$            \\	
Q\,1108$-$0747 & 	3.92	&	3.608	&	$20.37\pm0.07$	&	$-1.59\pm0.07$	&   $0.40\pm0.02$	& Si & 31  & $<14.0$    	      	&	$<14.2$		&	$<-5.59$		   & $h$            \\	
Q\,1111$-$1517 & 	3.37	&	3.266	&	$21.30\pm0.05$	&	$-1.65\pm0.11$	&   $0.34\pm0.09$	& Zn & 140 & $<13.1$    	      	&	$<13.4$		&	$<-7.37$		   & $h$            \\	
Q\,1117$-$1329 & 	3.96	&	3.351	&	$20.95\pm0.10$	&	$-1.41\pm0.11$	&   $0.23\pm0.06$	& Zn & 43  & $<14.0$        	        &	$<13.6$		&	$<-6.40$		   & $c$            \\	
Q\,1157$+$0128 & 	1.99	&	1.944	&	$21.80\pm0.10$	&	$-1.44\pm0.10$	&   $0.39\pm0.01$	& Zn & 89  & $<14.4$        	        &	$<14.5$		&	$<-6.65$		   & $b$            \\	
Q\,1209$+$0919 & 	3.30	&	2.584	&	$21.40\pm0.10$	&	$-1.01\pm0.10$	&   $0.53\pm0.03$	& Zn & 214 & $<14.9$          	        &	$<15.1$	     	&	$<-5.69$		   & $i$            \\  
Q\,1220$-$1800 & 	2.16	&	2.113	&	$20.12\pm0.07$	&	$-0.93\pm0.07$	&   $0.34\pm0.02$	& S  & 95  & $<13.8$	    	      	&	$<14.0$		&	$<-5.54$		   & $g,h$            \\  
Q\,1223$+$1753 & 	2.94	&	2.466	&	$21.40\pm0.10$	&	$-1.63\pm0.10$	&   $0.06\pm0.02$	& Zn & 91  & $<13.9$        	        &	$<14.1$		&	$<-6.69$		   & $b$            \\	
Q\,1232$+$0815 & 	2.57	&	2.338	&	$20.90\pm0.08$	&	$-1.43\pm0.08$	&   $0.45\pm0.01$	& S  & 85  & \multicolumn{2}{c}{$19.57^{+0.10}_{-0.10}$}        &	$-1.07^{+0.16}_{-0.17}$	   & ${b,l,r,s}$    \\	
Q\,1337$+$1121 & 	2.92	&	2.508	&	$20.12\pm0.05$	&	$-1.81\pm0.06$	&   $0.41\pm0.04$	& Si & 32  & $<13.8$       	       	&	$<14.2$		&	$<-5.42$		   & $c$            \\	
Q\,1337$+$1121 & 	2.92	&	2.796	&	$21.00\pm0.08$	&	$-1.86\pm0.09$	&   $0.31\pm0.04$	& Si & 42  & $<13.7$     	      	&	$<13.5$		&	$<-6.71$		   & $c$            \\	
Q\,1340$-$1340 & 	3.20	&	3.118	&	$20.05\pm0.08$	&	$-1.42\pm0.08$	&   $0.20\pm0.03$	& S  & 153 & $<12.5$    	      	&	$<12.9$		&	$<-6.62$		   & $h$            \\	
Q\,1354$-$1046 &        3.01    &       2.501   &       $20.44\pm0.05$  &       $-1.51\pm0.11$  &   $0.08\pm0.14$       & S  & 71  & $<14.7$                    &       $<14.5$         &       $<-5.18$                   & $h,j$            \\  
Q\,1354$-$1046 &        3.01    &       2.967   &       $20.80\pm0.10$  &       $-1.39\pm0.10$  &   $0.28\pm0.06$       & Si & 30  & $<13.1$                    &       $<13.5$         &       $<-6.75$                   & $h,j$            \\  
Q\,1418$-$0630 &        3.69    &       3.448   &       $20.50\pm0.10$  &       $-1.43\pm0.13$  &   $0.39\pm0.10$       & Si & 35  & $<14.0$                    &       $<14.2$         &       $<-5.69$                   & $h,j$            \\  
Q\,1441$+$2737 & 	4.42	&	4.224	&	$20.95\pm0.10$	&	$-0.63\pm0.10$	&   $0.49\pm0.03$	& Zn & 130 & \multicolumn{2}{c}{$18.29^{+0.07}_{-0.08}$}	&	$-2.36^{+0.17}_{-0.18}$	   & $i,t$            \\	
Q\,1444$+$0126 & 	2.21	&	2.087	&	$20.25\pm0.07$	&	$-0.80\pm0.09$	&   $0.95\pm0.07$	& Zn & 294 & \multicolumn{2}{c}{$18.16^{+0.14}_{-0.12}$}	&	$-1.80^{+0.21}_{-0.19}$	   & $c$            \\	
Q\,1451$+$1223 & 	3.25	&	2.469	&	$20.40\pm0.10$	&	$-2.27\pm0.13$	&   $0.25\pm0.08$	& Si & 35  & $<15.2$     	      	&	$<15.2$		&	$<-4.50$		   & $b$            \\  
Q\,1451$+$1223 & 	3.25	&	3.171	&	$20.20\pm0.20$	&	$-2.10\pm0.21$	&   $0.25\pm0.10$	& Si & 45  & $<13.6$     	      	&	$<13.6$		&	$<-5.80$		   & $b$            \\  
Q\,2059$-$3604 & 	3.09	&	2.507	&	$20.29\pm0.07$	&	$-1.85\pm0.20$	&   $0.44\pm0.19$	& S  & 25  & $<14.5$     	      	&	$<14.6$		&	$<-5.07$		   & $c$            \\	
Q\,2059$-$3604 & 	3.09	&	3.083	&	$20.98\pm0.08$	&	$-1.77\pm0.09$	&   $0.23\pm0.04$	& S  & 44  & $<13.4$     	      	&	$<13.5$		&	$<-6.85$		   & $b$            \\	
Q\,2116$-$3550 & 	2.34	&	1.996	&	$20.10\pm0.07$	&	$-0.34\pm0.11$	&   $0.54\pm0.09$	& Zn & 177 & $<14.5$          	        &	$<14.8$	     	&	$<-4.75$		   & $i$            \\  
Q\,2138$-$4427 & 	3.17	&	2.852	&	$20.98\pm0.05$	&	$-1.74\pm0.05$	&   $0.11\pm0.03$	& Zn & 58  & $<13.4$     	      	&	$<13.6$		&	$<-6.82$		   & $c$            \\	
Q\,2206$-$1958 & 	2.56	&	1.921	&	$20.67\pm0.05$	&	$-0.54\pm0.05$	&   $0.29\pm0.01$       & Zn & 136 & $<14.4$          	        &	$<14.7$	     	&	$<-5.44$		   & $i$            \\  
Q\,2206$-$1958 & 	2.56    &	2.076	&	$20.44\pm0.05$	&	$-2.32\pm0.05$	&   $0.28\pm0.02$	& Si & 20  & $<13.2$    	      	&	$<13.6$		&	$<-6.34$		   & $h$            \\  
Q\,2222$-$3939 & 	2.18	&	2.154	&	$20.85\pm0.10$	&	$-1.97\pm0.10$	&   $-0.04\pm0.04$	& S  & 21  & $<13.8$	    	      	&	$<13.8$		&	$<-6.35$		   & $g,h$            \\  
Q\,2228$-$3954 & 	2.21	&	2.095	&	$21.20\pm0.10$	&	$-1.36\pm0.12$	&   $0.17\pm0.07$	& Zn & 138 & $<13.8$    	      	&	$<14.0$		&	$<-6.59$		   & $g,h$            \\  
Q\,2230$+$0232 & 	2.15	&	1.864	&	$20.90\pm0.10$	&       $-0.81\pm0.10$	&   $0.24\pm0.01$	& S  & 148 & $<15.4$          	        &	$<15.4$	     	&	$<-4.80$		   & $i$            \\  
Q\,2243$-$6031 & 	3.01	&	2.331	&	$20.65\pm0.05$	&	$-0.85\pm0.05$	&   $0.35\pm0.02$	& Zn & 173 & $<13.8$          	        &    	$<13.9$	     	&	$<-6.15$		   & $i$            \\  
Q\,2311$-$3721 &        2.48    &       2.182   &       $20.55\pm0.07$  &       $-1.57\pm0.10$  &   $0.26\pm0.05$       & Si & 77  & $<14.7$                    &       $<15.0$         &       $<-5.00$                   & $h,j$            \\  
Q\,2318$-$1107 & 	2.96	&	1.989	&	$20.68\pm0.05$	&	$-0.85\pm0.06$	&   $0.42\pm0.03$	& Zn & 207 & \multicolumn{2}{c}{$15.49^{+0.03}_{-0.03}$}	&	$-4.89^{+0.08}_{-0.08}$	   & $f,g$            \\  
Q\,2332$-$0924 & 	3.32	&	3.057	&	$20.50\pm0.07$	&	$-1.33\pm0.08$	&   $0.30\pm0.03$	& S  & 111 & $<13.2$     	      	&	$<13.5$		&	$<-6.45$		   & $c$            \\	
Q\,2343$+$1232 & 	2.51	&	2.431	&	$20.40\pm0.07$	&	$-0.87\pm0.10$	&   $0.51\pm0.07$       & Zn & 289 & \multicolumn{2}{c}{$13.69^{+0.09}_{-0.09}$}	&	$-6.41^{+0.16}_{-0.16}$	   & ${f,i}$        \\	
Q\,2344$+$1229 & 	2.76	&	2.538	&	$20.50\pm0.10$	&	$-1.81\pm0.10$	&   $0.18\pm0.01$	& Si & 69  & $<13.4$    	      	&	$<13.4$		&	$<-6.40$		   & $h$            \\	
Q\,2348$-$0108 & 	3.01	&	2.426	&	$20.50\pm0.10$	&	$-0.62\pm0.10$	&   $0.55\pm0.02$       & S  & 248 & \multicolumn{2}{c}{$18.52^{+0.29}_{-0.49}$}	&	$-1.69^{+0.38}_{-0.58}$	   & ${i,u}$        \\  
Q\,2348$-$0108 & 	3.01	&	2.615	&	$21.30\pm0.08$	&	$-2.02\pm0.08$	&   $0.13\pm0.01$	& Si & 95  & $<13.1$    	      	&	$<13.6$		&	$<-7.20$		   & $h,u$            \\  
Q\,2348$-$1444 & 	2.94	&	2.279	&	$20.63\pm0.05$	&	$-2.10\pm0.08$	&   $0.25\pm0.10$	& S  & 55  & $<13.6$    	      	&	$<13.6$		&	$<-6.38$		   & $h$            \\  

\end{longtable}

\footnotesize
\noindent
$^a$ \citet{Levshakov00,Levshakov01},  
$^b$ \citet{Petitjean00}, 
$^c$ \citet{Ledoux03},  
$^d$ \citet{Ge97}, 
$^e$ \citet{Petitjean02}, 
$^f$ \citet{Noterdaeme07lf}, 
$^g$ \citet{Smette05}, 
$^h$ this work, 
$^i$ \citet{Petitjean06},
$^j$ \citet{Akerman05}, 
$^k$ \citet{Levshakov02}, 
$^l$ \citet{Srianand05},
$^m$ \citet{Heinmuller06},
$^n$ \citet{Levshakov85},
$^o$ \citet{Foltz88},
$^p$ \citet{Srianand98},
$^q$ \citet{Ledoux02}, 
$^r$ \citet{Srianand00},
$^s$ \citet{Ge01},
$^t$ \citet{Ledoux06b},
$^u$ \citet{Noterdaeme07}.\\
$^\dagger$ $N$(\ion{Fe}{ii}) from \citet{Prochaska07b}.\\
$^\ast$ Metal column densities from \citet{Prochaska07b}. Only an upper limit on $N$(\FeII) could be measured \citep{Prochaska03}. 
The abundance of iron is therefore estimated 
through that of nickel, assuming [Fe/H]~=~[Ni/H$]-0.1$~dex.
\\
\normalsize
}
